\documentclass[reprint, superscriptaddress, amsmath,amssymb, aps,floatfix, prb]{revtex4-2}
\usepackage{hyperref}
\usepackage{graphicx} % Required for inserting images
\usepackage{geometry}
\usepackage{color}

\newcommand{\bk}{\mathbf{k}}
\newcommand{\bq}{\mathbf{q}}
\newcommand{\bQ}{\mathbf{Q}}

\newcommand{\intd}{\mathrm{d}}

\newcommand{\imag}{\textrm{Im}}
\newcommand{\QPI}{\mathcal{I}}
\newcommand{\SOL}{\mathcal{S}}
\newcommand{\FeSeS}{\mathrm{FeSe}_{1-x}\mathrm{S}_{x}}

\newcommand{\fesesonenine}{$\textrm{Fe}\textrm{Se}_{0.81}\textrm{S}_{0.19}$}

\date{\today}

\begin{document}
\title{Intermediate band analysis in Green’s functions calculations of quasiparticle interference}

\author{Xinze Yang}
\affiliation{\footnotesize \mbox{Department of Physics, Yale University, New Haven, Connecticut 06520, USA}}
\affiliation{\footnotesize \mbox{Energy Sciences Institute, Yale University, West Haven, Connecticut 06516, USA}}

\author{Alexander F.\,Kemper}
\affiliation{\footnotesize Department of Physics, North Carolina State University, Raleigh, NC 27695, USA}

\author{Adrian Gozar}
\affiliation{\footnotesize \mbox{Department of Physics, Fairfield University, Fairfield, CT 06824}}
\affiliation{\footnotesize \mbox{Department of Physics, Yale University, New Haven, Connecticut 06520, USA}}
\affiliation{\footnotesize \mbox{Energy Sciences Institute, Yale University, West Haven, Connecticut 06516, USA}}

\author{Eduardo H.\,da Silva Neto}
\email[Corresponding Author: ]{eduardo.dasilvaneto@yale.edu}
\affiliation{\footnotesize \mbox{Department of Physics, Yale University, New Haven, Connecticut 06520, USA}}
\affiliation{\footnotesize \mbox{Energy Sciences Institute, Yale University, West Haven, Connecticut 06516, USA}}
\affiliation{\footnotesize \mbox{Department of Applied Physics, Yale University, New Haven, Connecticut 06520, USA}}

\begin{abstract}
The measurement of quasiparticle scattering patterns on material surfaces using scanning tunneling microscopy (STM) is now an established technique for accessing the momentum-resolved electronic band structure of solids. However, since these quasiparticle interference (QPI) patterns reflect spatial variations related to differences in the band momenta rather than the momenta themselves, their interpretation often relies on comparisons with simple geometrical models such as the joint density of states (JDOS) or with the convolution of Green’s functions. 
In this paper, we highlight non-intuitive differences between Green’s function and JDOS results. To understand the origin of these discrepancies, we analyze the convolution of Green's functions using the Feynman parametrization technique and introduce a framework that we call the intermediate band analysis. This approach allows us to derive simple selection rules for interband QPI, based on electron group velocities. Connecting the intermediate band analysis with the experiment, we consider experimental Bogoliubov QPI patterns measured for $\FeSeS$, which were recently used to demonstrate a highly anisotropic superconducting gap, indicating superconductivity mediated by nematic fluctuations \cite{Nag2024}. The calculated Green's functions convolutions reproduce the particle-hole asymmetry in the intensity of QPI patterns across the Fermi level observed in experiments. Finally, we demonstrate the utility of intermediate band analysis in tracing the origin of this asymmetry to a coherence factor effect of the superconducting state.

\end{abstract}
\maketitle
\section{Introduction}
Quasiparticle interference (QPI) occurs when an electron scatters off an impurity, creating interference patterns that result in spatial oscillations in the local density of states (LDOS). The wave vector $\bq$ of this oscillation corresponds to the difference between the initial and final electronic states' Bloch vectors, \textit{i.e}, $\bq=\bk_f - \bk_i$. This $\bq$ vector can be measured using spectroscopic mapping techniques with a scanning tunneling microscope (STM). Over the past two decades, QPI measurements have become an established method for accessing the electronic band structure of solids \cite{Hoffman2002a,Hoffman2002b,Capriotti2003}.
However, a significant challenge in relating the QPI patterns to a complicated multi-band system arises from the fact that, in this technique, $\bk$ is only accessed indirectly via $\bq$. To infer the $\bk$ states from $\bq$ measurements, it is common to compare the experimental data to calculations. There are two commonly used calculation methods: one calculates the convolution of unperturbed Green's functions and the other calculates the joint density of states (JDOS) \cite{Wang2003,Capriotti2003,Balatsky2006,Hirschfeld2015}.

Under the JDOS method, at a given energy $\omega$, QPI patterns are determined by the available points on the contour of constant energy (CCE), \textit{i.e.}, $\bq(\omega) = \bk_f(\omega) - \bk_i(\omega)$. This approximation turns the issue of interpreting QPI patterns into a geometrical exercise, which is convenient for experimentalists who need to quickly observe and interpret STM data during experiments. A celebrated achievement of the JDOS approach is the octet model for anisotropic superconductors. This model explained the most salient features of the QPI observed in cuprate superconductors \cite{Hoffman2002a, Hoffman2002b, Wang2003} and has been successful in resolving the momentum structure of superconducting gaps in various quantum materials \cite{Hoffman2002b, Allan2013, Sprau2017, Nag2024}. 

Those early results eventually led to a simple protocol. Given a measured QPI pattern, the first step is to try to identify the geometrical relations $\bq(\omega) = \bk_f(\omega) - \bk_i(\omega)$. However, in situations where this is impractical, such as when there are too many overlapping features in a multi-band system, one typically resorts to a perturbative calculation relying on the Green's function. This method usually starts by determining a reasonable $\bk$-space band structure of the system (\textit{e.g.}, from angle-resolved photoemission spectroscopy measurements or from theoretical calculations) and then perturbatively calculating the LDOS using Green's functions. As will be discussed later, this method can be simplified to the calculation of the convolution of Green's functions under several reasonable approximations, which have been commonly applied in previous research works \cite{Wang2003,Capriotti2003,Balatsky2006,Hirschfeld2015, Walker2023}. An underlying assumption of this protocol is that the calculation results using Green’s functions serve as a refinement of the JDOS interpretation. In other words, it assumes that both methods will qualitatively identify the same $\bq$-space locations for the poles of the response function, but with the Green’s function approach providing a more precise representation of the QPI feature weights. As will be shown in this paper, this point of view is not true and QPI calculated through JDOS and Green's function can be fundamentally different. 

Motivated by recent experimental studies of the $\FeSeS$ Fe-based superconductors by some of the present authors, which report detailed QPI measurements of the normal and superconducting states of \fesesonenine{} \cite{Walker2023, Nag2024}, and considering the numerous STM studies that compare measurements with QPI calculated through Green’s function \cite{Capriotti2003, Wang2003, Hirschfeld2015, Rhodes2019}, we initiated an investigation to better understand the differences between JDOS and Green’s function convolution, along with the physical origins of these discrepancies. We first discuss the general formalism for the two methods in Sec.\,\ref{sec:JDOS_Green}. Then, in Sec.\,\ref{sec:II}, we show that even in the simplest multi-band system, composed of only two parabolic bands, a qualitative discrepancy exists between the JDOS and Green's function convolution. While the JDOS calculation yields two interband scattering processes, the Green's function convolution yields only one, depending on the relative sign between the masses of the two bands. We show analytically how this can be understood in terms of a group velocity rule for the one-dimensional case. In Sec.\,\ref{sec:intermediate_bands}, we explore the case more relevant to STM experiments, the two-dimensional case, using a Feynman parametrization technique to restructure the Green's function QPI problem from a first order two-pole integral into a second order single-pole integral. In this new representation, the QPI signal is determined by the single poles of new bands, which we call intermediate bands, and the final QPI signal is obtained by summing over all possible intermediate bands. Analyzing the interband QPI problem through the lens of these intermediate bands allows us to extend the group velocity selection rule from the 1D to the 2D case. In Sec.\,\ref{sec:BQPI}, we extend the intermediate band analysis to the case of an anisotropic superconductor. Using the band and gap parameters that describe the superconducting Bogoliubov QPI (BQPI) in \fesesonenine{}, we find a particle-hole asymmetry in the intensity of BQPI features that is similar to experimental observations. Analyzing the problem in terms of the intermediate bands, we attribute the source of this asymmetry to a coherence factor effect, determined by the combinations of Bogoliubov coefficients, $u_\bk$ and $v_\bk$. We present our conclusions in Sec.\,\ref{sec:conclusions}.

\section{JDOS AND GREEN'S FUNCTION FORMALISM\label{sec:JDOS_Green}}
We begin with a brief review of the Green’s function and JDOS methods for QPI calculation\cite{Bruus2004}. Under the perturbation of impurities, the Green's function can be calculated using the the $T$-matrix approach:
\begin{equation}
\label{eq: T Matrix Approximation}
G(\bk, \bk^{\prime}, \omega) = G_0(\bk, \omega) + G_0(\bk, \omega) T_{\bk, \bk^{\prime}}(\omega) G_0(\bk^{\prime}, \omega)
\end{equation}
where $G_0$ is the Green's function of the impurity-free material and $T_{\bk, \bk^{\prime}}(\omega)$ is the $T$-matrix of the impurity potential. 
% Once obtain the Green's function, the LDOS can be firstly calculated in real space:
% \begin{equation}
% A(\br, \omega) = -\frac{1}{\pi}\imag G(\br, \br,\omega)
% \end{equation}
% where $G(\br, \br,\omega)$ is the Green's function represented in real space. 
Typically, QPI data is analyzed in the $\bq$ domain, with the absolute value of the Fourier transform of the STM real-space data presented. Under this representation, the change of LDOS due to impurity perturbation (denoted as $\delta A(\bq, \omega)$) is given by:
\begin{equation}
\label{eq: Def Change of LDOS in q Space}
\delta A(\bq, \omega) = -\frac{1}{\pi}\imag \int \frac{\intd^2\bk}{(2 \pi)^2}G_0(\bk, \omega)T_{\bk, \bk-\bq}(\omega)G_0(\bk-\bq, \omega)
\end{equation}

Experimentally, STM measurements do not reveal the specific form of the impurity potential, making $T_{\bk, \bk-\bq}(\omega)$ unknown. Consequently, it is usually disregarded from the problem, and instead, a convolution of the Green’s function in $\bk$-space is calculated, denoted by
\begin{equation}
\label{eq: QPI Convolution Definition}
\QPI(\bq, \omega) = \imag \int \frac{\intd^2\bk}{(2 \pi)^2} G_{0}(\bk, \omega) G_{0}(\bk - \bq, \omega).
\end{equation}
Formally, this form can be obtained from either of two distinct approximations. One is to assume the impurity potential is a $\delta$-function, \textit{i.e.} the effect of the impurity is absolutely localized. Under this assumption, the $T$ matrix is independent of $\bk$ and the LDOS in $\bq$ space is proportional to the imaginary part of the convolution of the Green's function, $\QPI(\bq, \omega)$. 
Alternatively, even if the impurity potential has finite effective radius in real space, the $T$ matrix can be approximated to first order as $V(\bq)$. Under this approximation
\begin{equation}
\delta A(\bq, \omega) \approx -\frac{V(\bq)}{\pi} \imag \int \frac{\intd^2 \bk}{(2\pi)^2} G_0(\bk, \omega)G_0(\bk-\bq, \omega)
\end{equation}
Since $V(\bq)$ is just a multiplier in the whole expression, the LDOS modulations are still represented by the convolution term \cite{Capriotti2003}. Therefore, $\QPI(\bq, \omega)$ (Eq.\,(\ref{eq: QPI Convolution Definition})) is often used to simulate QPI patterns for comparison to experimental data. Throughout this paper, this function, $\QPI(\bq, \omega)$, is referred to as \textit{the QPI response function} \cite{Capriotti2003}.

Ultimately, we would like to understand the origin of experimental QPI features in terms of $\bk$ and $\bk^{\prime}=\bk-\bq$. However, visualizing the convolution of two complex functions in $\QPI(\bq, \omega)$ is difficult, making it formally impossible to intuitively predict $\bk$ and $\bk^{\prime}$ from the form of Eq.\,\ref{eq: QPI Convolution Definition}. Still, as a first approximation, one might expect the structure of $\QPI(\bq, \omega)$ to show peaks for $\bq$ vectors that allow the poles $G_0(\bk,\omega)$ to overlap with poles in $G_0(\bk-\bq,\omega)$. This geometrical criterion is captured by the joint density of states (JDOS), $J(\bq, \omega)$, which is the auto-convolution of the spectral function
\begin{equation}
\label{eq: JDOS Definition}
J(\bq, \omega) = \int\intd\bk A_0(\bk, \omega) A_0(\bk - \bq, \omega)
\end{equation}

The QPI response function, $\QPI(\bq, \omega)$, is often regarded as a more accurate representation of the relative intensities of features in $\bq$ space, as the JDOS method does not account for the constructive and destructive interference effects inherent to complex numbers. Nevertheless, it is also frequently stated that the $\bq$ locations of features and peaks are the same in both $\QPI(\bq, \omega)$ and $J(\bq, \omega)$.
Building on these intuitions, a common approach when dealing with multiband systems is to numerically evaluate Eq.\,\ref{eq: QPI Convolution Definition} and then attempt to geometrically relate the features in $\QPI(\bq, \omega)$ to the corresponding $\bk$ and $\bk^{\prime}$. However, this approach and its underlying intuition can be problematic. As demonstrated through the examples in Sec.,\ref{sec:II}, there are generic cases where JDOS features are entirely absent in $\QPI(\bq, \omega)$, highlighting the limitations of this methodology.

\subsection{QPI from a single band}

\begin{figure}[h]
    \centering
    \includegraphics[width=1.0\linewidth]{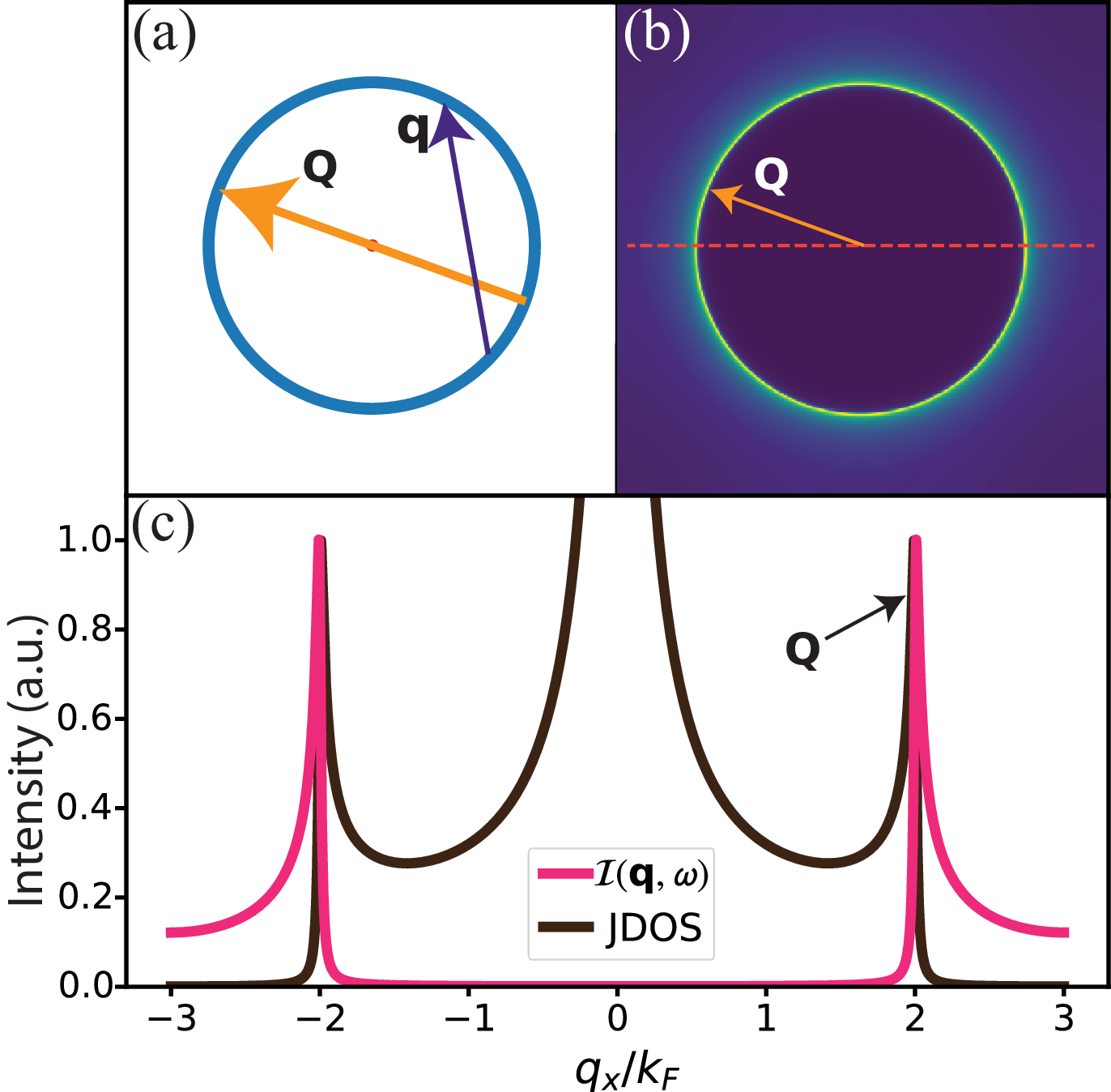}
    \caption{(a) Fermi circle of the quadratic band. $\bQ$ and $\bq$ are two vectors have non-zero JDOS. The image has size $3k_{F}\times 3k_{F}$. (b) Numerically calculated  QPI response function where only $\bQ$ have finite QPI intensity. The image has size $6k_{F}\times 6k_{F}$. (c) The line cut of QPI calculated by JDOS and QPI response function along $q_y=0$ (red dashed line in (b)).}
    \label{fig: single band}
\end{figure}

We begin to explore the differences between $\QPI(\bq, \omega)$ and $J(\bq, \omega)$ with the simplest possible example.  Consider an isotropic 2D quadratic band structure with a circular Fermi surface with Fermi momentum $k_F$. Since any $\bq$ vector with length smaller than $2k_{F}$ connects two points on the Fermi surface (Fig.\;\ref{fig: single band}(a)), the JDOS predicts non-zero QPI intensities for all $0<|\bq|<2k_F$ (Fig.\;\ref{fig: single band}(c)). At the two extremes, a high intensity is expected at $\bq=0$, where the Fermi circle overlaps with itself, and at $\bq=2 k_F=\bQ$, where two circles are externally tangent to each other. However, the QPI response function, either through the analytical \cite{Capriotti2003} or numerical (Fig.\,\ref{fig: single band}(b)) evaluation of Eq.\,(\ref{eq: QPI Convolution Definition}) (the red curve in Fig.\,\ref{fig: single band}), reveals zero QPI intensity at all points within the circle of radius $2k_F$ except at the boundary and the absence of a $\bq=0$ peak, Figs.\,\ref{fig: single band}(b,c). 
This example illustrates why the $\QPI(\bq, \omega)$ is often interpreted through the lens of the JDOS: both calculations display peaks at $\bQ$. However, the discrepancies shown in Fig.\,\ref{fig: single band} reveal that the complex nature of the Green’s function introduces nontrivial phenomena in $\QPI(\bq,\omega)$ that are absent in the JDOS, which we explore in the next section.
\section{INTERBAND QPI BETWEEN TWO QUADRATIC BANDS: A PUZZLE \label{sec:II}}
\subsection{Parabolic bands in 2D \label{ssec:Puzzle}}
\begin{figure}[htbp]
    \includegraphics[width=1\linewidth]{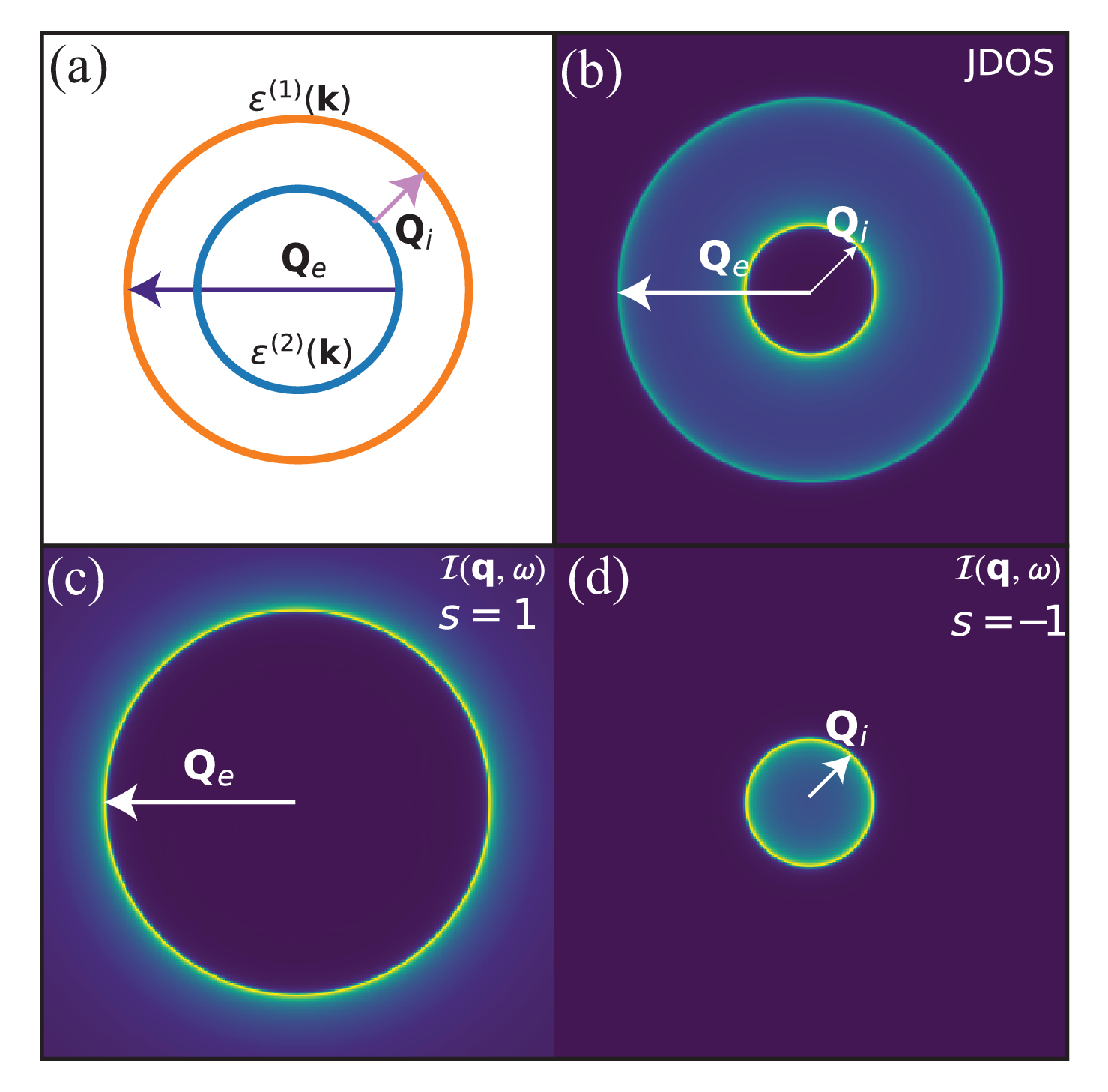}
    \caption{(a) Fermi surfaces of the two quadratic bands given by Eqs.\,(\ref{eq: Band1}) and (\ref{eq: Band2}). Two bands here are related by $\mu_2 = 0.625\mu_1$ and $m_2 = 0.4m_1$. The size of the frame is $3 k_{F}^{(1)}\times 3 k_{F}^{(1)}$. (b) Numerical JDOS calculation results for both $s=1$ and $s=-1$ (JDOS result are the same for these two cases). (c) and (d) Numerical QPI response function results for the system illustrated by (a). Images (b) to (d) have size $4 k_{F}^{(1)}\times 4 k_{F}^{(1)}$.}
    \label{fig: Puzzle2D}
\end{figure}
% \begin{figure}[htbp]
%     \includegraphics[width=1\linewidth]{Fig_Puzzle2D.eps}
%     \caption{(a) Fermi surfaces of two bands given by Eq.\,(\ref{eq: Band1}) and (\ref{eq: Band2}). Two bands here are related by $\mu_2 = 0.875\mu_1$ and $m_2 = 0.4m_1$. The size of the frame is $3 k_{F}^{(1)}\times 3 k_{F}^{(1)}$. (b): Numerical JDOS calculation results for both $s=1$ and $s=-1$ (JDOS result are the same for these two cases). (c) and (d): Numerical QPI interference response function results for the system illustrated by (a). Images (b) to (d) have size $4 k_{F}^{(1)}\times 4 k_{F}^{(1)}$.}
%     \label{fig: Puzzle2D}
% \end{figure}

We now focus on the interband scattering problem. For this, we analyze a simple band structure composed of two quadratic bands $\epsilon^{(1)}(\bk)$ and $\epsilon^{(2)}(\bk)$, 
\begin{align}
    \label{eq: Band1} \epsilon^{(1)}(\bk) = & \frac{\bk^2}{2m_1} + \mu_1\\
    \label{eq: Band2} \epsilon^{(2)}(\bk) = & s\frac{\bk^2}{2m_2} + s\mu_2
\end{align}
where the first band has positive effective mass (particle band) and the effective mass sign of the second, denoted as $s\in\{-1, 1\}$, is free to change (hole or particle band). 
%For convenience, the two bands are adjusted to keep the Fermi level at zero energy. 
We will only consider the interband QPI at a single energy level, which for simplicity we adjust to be the Fermi level, but the results are easily generalized to any contour of constant energy. Similar to the single band QPI discussed before (Fig.\,\ref{fig: single band}), interband QPI calculated from the JDOS results in circles with radii 
$$|\bQ_{e}| = k^{(1)}_{F} + k^{(2)}_{F}$$ 
and 
$$|\bQ_{i}| = |k^{(1)}_{F} - k^{(2)}_{F}|$$
where the subscripts $e$ and $i$ denote situations when the two Fermi circles are externally or internally tangent with each other and $k^{(1,2)}_{F}$ are Fermi momenta of two bands (Fig.\,\ref{fig: Puzzle2D}(a)). Within the JDOS approximation, the existence of these two peaks is not affected by the sign factor since changing $s$ from $-1$ to $1$ leaves the Fermi surface geometry invariant. As a result, circles with radii $|\bQ_e|$ and $|\bQ_i|$ appear (Fig.\,\ref{fig: Puzzle2D}(b)) in both particle-particle ($s=1$) and particle-hole systems ($s=-1$).

The QPI response function, however, shows qualitatively different QPI patterns. As shown in Fig.\,\ref{fig: Puzzle2D}(c), the $|\bQ_{e}|$ feature only appears when $s = 1$ (particle-particle system), while the feature corresponding to $|\bQ_{i}|$ only appears when $s=-1$. These numerical results suggest the existence of additional selection rules, where phenomenologically only $\bq$ vectors linking points with opposite group velocity can have a peak in the interband QPI response function. 

It is worth noting that this selection rule seems to hold experimentally. Perhaps the clearest example we found in the literature comes from observation in 2D $\mathrm{ErSi_2}$ on $\mathrm{Si}(111)$ \cite{Simon2011}. Its band structure is composed of one circular hole pocket at the $\Gamma$ point and six equivalent elliptical particle pockets at $M$ points. In the experiment, it was observed that the QPI pattern between the bands between these two pockets takes the form of a `butterfly' shape, in matter consistent with the opposite group velocity prediction, as shown in Fig.\,\ref{fig: PosNegInHomo} and discussed in Appendix\,\ref{Appendix: Saddle Point}.
\begin{figure}[htbp]
    \includegraphics[width=1\linewidth]{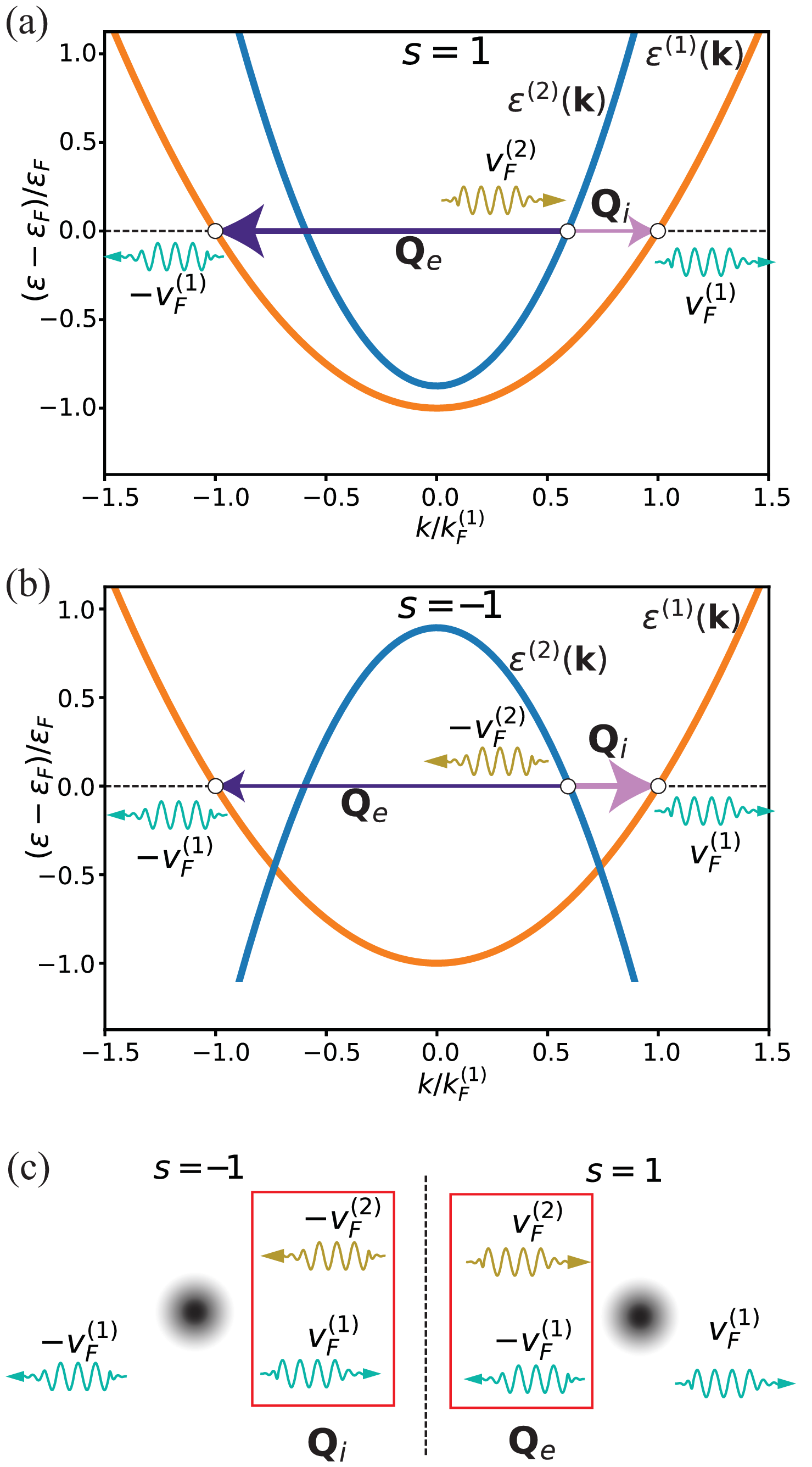}
    \caption{(a) and (b): Illustrations of the two bands system in 1D with $s=1$ (a) and $s=-1$ (b). Two bands plotted here are related by $\mu_2 = 0.875\mu_1$ and $m_2 = 0.4m_1$. C: A illustration of the wave scattering picture.}
    \label{fig: Puzzle1D}
\end{figure}

\subsection{Opposite group velocity rule in the 1D Case}
An intuitive explanation for the opposite group velocity selection rule can be obtained in 1D case. For an arbitrary 1D band $\epsilon(k)$ with group velocity $v_0$ at point $k_0$, the linearized Green's functions at 
%the energy level $\epsilon(k_0)$
the Fermi level, denoted as $G_{k_0}(k)$, is given by:
\begin{equation}
\label{eq: 1D Linear Green's Function}
G_{k_0}(k) = \frac{1}{-v_0(k-k_0) + i \eta^{+}}
\end{equation}
whose real space representations is
\begin{equation}
\label{eq: 1D Linear Green's Function Real Representation}
G_{k_0}(x) = -\frac{i}{|v_0|}\theta \bigg(\frac{x}{v_0}\bigg)e^{i k_0 x}
\end{equation}
Note that $k$ in the Eq.\,(\ref{eq: 1D Linear Green's Function}) takes on a similar role as the $\omega$ in a causal Green's function. Thus this linearized Green's function is \textit{spatially `causal'}, with its non-zero domain determined by the $\theta$ function.

Now consider the interband QPI between $\epsilon^{(1)}(k)$ and $\epsilon^{(2)}(k)$. Fourier transforming Eq.\,(\ref{eq: QPI Convolution Definition}) to real space for the 1D case, the QPI intensity becomes:
\begin{equation}
\label{eq: QPI Real Representation}
\QPI(x, \omega=0) = \imag\{G^{(1)}(x) G^{(2)}(-x)\}
\end{equation}
where $G^{(1, 2)}(x)$ are Green's functions of bands $\epsilon^{(1, 2)}(k)$ in real space. For a specific scattering process from $k_1$ to $k_2$, Eqs.\,(\ref{eq: 1D Linear Green's Function Real Representation}) and (\ref{eq: QPI Real Representation}) give the real space QPI intensity to be
\begin{equation}
\label{eq: General 1D Scattering Intensity}
\imag\bigg\{-\frac{1}{|v_1 v_2|}\theta\bigg(\frac{x}{v_1}\bigg)\theta\bigg(-\frac{x}{v_2}\bigg)e^{i(k_1-k_2)x}\bigg\}
\end{equation}
Here $v_{1,2}$ are group velocities of the band 1 and 2 at $k_{1,2}$ relatively. The product of two $\theta$ functions in Eq.\,(\ref{eq: General 1D Scattering Intensity}) forbids any scattering intensity when $v_1$ and $v_2$ have same sign.

Now we apply this result to the quadratic bands system. Figures \ref{fig: Puzzle1D}(a) and (b) depict the 1D particle-particle and particle-hole scenarios, where the vectors $\bQ_e$ and $\bQ_i$, marked in the figures, are the only possible interband scattering processes. When $s=1$, Fig.\,\ref{fig: Puzzle1D}(a), the QPI intensity at $\bQ_i$ is suppressed since it connects two points with parallel group velocities. On the other hand, QPI is finite for $\bQ_e$ since this $\bQ$ vector connect points with anitparallel group velocities, causing non-zero regions of two theta functions in Eq.\,(\ref{eq: QPI Real Representation}) overlap. 
In contrast, when $s=-1$, QPI is suppressed at $\bQ_e$ but exists for $\bQ_i$. From the point of view of wave scattering, this opposite group velocity rule indicates that the incoming and outgoing waves must spatially overlap to interfere and form a measurable standing wave, as illustrated in Fig.\,\ref{fig: Puzzle1D}(c). For example, consider the $s=-1$ case depicted in Fig.\,\ref{fig: Puzzle1D}(b). The allowed wave vector $\bQ_i$ corresponds to scattering between two states on the right side of the impurity. Likewise the wave vector $\bQ_e$ is forbidden because it involves left movers on opposite sides of the impurity.

Unfortunately, a similar analysis cannot be adapted directly to 2D case, which actually gathers most experimental interest. This is because in 2D, there are infinite possible directions for scattering, leading to more complicated interference relation between the incoming and outgoing wave. To overcome this difficulty, in next section, we will apply Feynman parameterization to evaluate Eq.\,(\ref{eq: QPI Convolution Definition}). In the 2D case, we show that $\bq$ vectors connecting two points with opposite group velocities exhibit diverging QPI intensity, while $\bq$ vectors linking points with parallel group velocities produce no signal.

\section{INTERMEDIATE BANDS EVOLUTION FOR 2D QUADRATIC BANDS \label{sec:intermediate_bands}}
\subsection{Intermediate Bands}

To expand the 1D analysis in Sec.\,\ref{sec:II} to the 2D case, we introduce an approach to evaluate and visualize interband scattering, leveraging the similarity between $\QPI(\bq, \omega)$ and loop integrals in quantum field theory. The key element is the use of the Feynman parametrization technique, which modifies the convolution in Eq.\,(\ref{eq: QPI Convolution Definition}) from a two-pole integral to a single-pole integral \cite{Schwartz2014}. As Feynman observed:
\begin{equation}
\frac{1}{AB} = \int_{0}^{1}\frac{\intd t}{(t A + (1 - t) B)^2} 
\end{equation}
Applying this to the QPI response function defined by Eq.\,(\ref{eq: QPI Convolution Definition}) gives
\begin{align}
\label{eq: Def QPI Convolution}
\QPI(\bq, \omega) & = \imag \int\frac{\intd^2\bk/(2\pi)^2 }{(z-\epsilon^{(1)}(\bk))(z-\epsilon^{(2)}(\bk-\bq))}\\
\label{eq: Def QPI Self-Overlapping}
&= \imag \int_{0}^{1}\int\frac{\intd t\intd^2\bk/(2\pi)^2 }{(z-[t \epsilon^{(1)}(\bk) + (1-t)\epsilon^{(2)}(\bk-\bq)])^2} 
\end{align}
Here $z$ is a complex number $\omega + i \eta^+$ where $\eta^+$ is a small positive number tending to zero at the end of the calculation. Defining a new band structure $\epsilon_{t, \bq}^{*}(\bk)$ in $\bk$ space with parameters $t$ and $\bq$:
\begin{equation}
\label{eq: Def Intermediate Band}
\epsilon_{t, \bq}^{*}(\bk) = t \epsilon^{(1)}(\bk) + (1-t)\epsilon^{(2)}(\bk-\bq)
\end{equation}
and switching the order of integration, Eq.\,(\ref{eq: Def QPI Self-Overlapping}) becomes:
\begin{equation}
\label{eq: Def QPI Intermediate Band}
\QPI(\omega, \bq) = \imag \int_{0}^{1} \intd t \int\frac{1}{(\omega- \epsilon_{t, \bq}^{*}(\bk) + i \eta^{+})^2}\frac{\intd\bk^2}{(2\pi)^2}
\end{equation}
Equation (\ref{eq: Def QPI Intermediate Band}) now has a single second order pole structure determined by $\epsilon_{t, \bq}^{*}(\bk)$, which is a weighted average of bands $\epsilon^{(1)}(\bk)$ and $\epsilon^{(2)}(\bk - \bq)$ at a given $t$. Varying $t$ evolves the band $\epsilon_{ x, \bq}^{*}(\bk)$ continuously from $\epsilon^{(2)}(\bk - \bq)$ to $\epsilon^{(1)}(\bk)$. Therefore, we refer to $\epsilon_{t, \bq}^{*}(\bk)$ as the \textit{intermediate band}. 

We now turn to the $\bk$ integral appearing in Eq.\,(\ref{eq: Def QPI Intermediate Band}). For a generic band $\epsilon(\bk)$, we need to evaluate an integral of the form
\begin{equation}
\SOL_{\omega}[\epsilon] = \imag \int \frac{1}{(\omega- \epsilon(\bk) + i \eta^{+})^2}\frac{\intd^2\bk}{(2\pi)^2}
\end{equation}
This integral is simply the derivative of the density of states (d-DOS) of the band $\epsilon(\bk)$ (Appendix \ref{Appendix: Ddos Proof})
\begin{equation}
\label{eq:SOL and DOS relation}
\SOL_{\omega}[\epsilon] = \pi\frac{\intd g(\omega)}{\intd \omega}
\end{equation}
Denoting the DOS of the intermediate band $\epsilon^*_{t, \bq}(\bk)$ as $g^*_{t, \bq}(\omega)$, QPI described by Eq.\,(\ref{eq: Def QPI Convolution}) can be written as
\begin{equation}
\label{eq: QPI Integral of DDOS}
\QPI(\bq, \omega) = \pi\int_{0}^{1}\frac{\intd g^*_{t, \bq}(\omega)}{\intd \omega}\intd t
\end{equation}
This equation illustrates that the QPI intensity at a point $(\bq, \omega)$ is the integral of the d-DOS of all intermediate bands. Therefore, we have reformulated the problem of analyzing QPI response function from Eq.\,(\ref{eq: QPI Convolution Definition}) into an analysis of the evolution of intermediate bands, and their DOS, with $t$.

\begin{figure}[htbp]
    \includegraphics[width=1\linewidth]{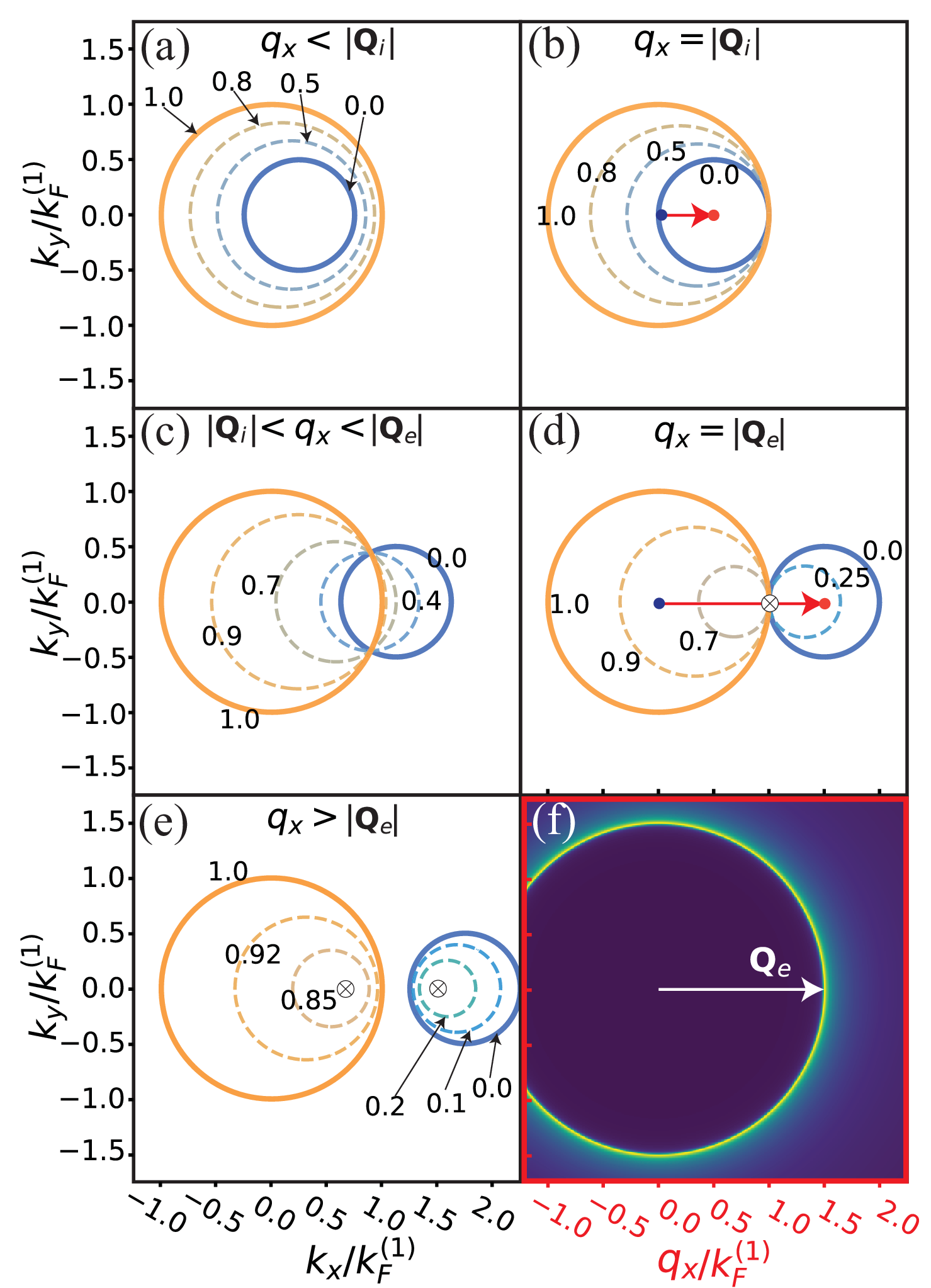}
    \caption{(a)-(e) Evolution of Fermi surfaces of intermediate bands with $q_x$ values in different ranges for the particle-particle system given in Fig.\,\ref{fig: Puzzle2D} ($s=1$). Numbers marked around Fermi surfaces are corresponding $t$ values. ETPs are marked by the symbol $\otimes$ in the figure. Two red arrows in (b) and (d) marks the $\bQ_i$ and $\bQ_e$. (f) Numerical results of QPI response function. (a)-(e) are plots in $\bk$ space (black frame) and (f) is a plot in $\bq$ space (red frame)}
    \label{fig: PositiveMassEvolution}
\end{figure}

\subsection{Intermediate Bands Analysis of Interband Scattering: Particle-Particle System}
We now use the intermediate bands reformulation to analyze the interband scattering discussed in Sec.\,\ref{ssec:Puzzle}. First note that the DOS of the 2D electron gas has the special property that its DOS is a step function jumping at the band minimum/maximum. Given an arbitrary 2D quadratic band
\begin{equation}
\epsilon(k_x, k_y) = \frac{k_x^2}{2m_x} + \frac{k_y^2}{2m_y} + \mu
\end{equation}
with the band minimum at $\mu$, the d-DOS is given by:
\begin{equation}
\label{eq:Quadratic DDOS}
\frac{\intd g(\omega)}{\intd \omega} = \frac{\sqrt{m_xm_y}}{2\pi}\delta(\omega-\mu) 
\end{equation}
Thus the $\bq$ values where QPI signal is expected from the QPI response function are determined by intermediate bands with minimum or maximum touching the energy level $\omega$. Equivalently, in $\bk$ space, a band minimum or maximum touching  $\omega$ is equivalent to the contour of constant energy shrinking to a single point. We refer to this as the extremum touching point (ETP) and its existence within $0<t<1$ will be important to determine whether QPI exists for a given $\bq$.

Figure \ref{fig: PositiveMassEvolution} shows the intermediate band evolution for the 2D quadratic system considered in Sec.\,II, for various values of $\bq = (q_x, 0)$ in the particle-particle case ($s=1$). When $q_x \leq |\bQ_i|$, Figs.\,\ref{fig: PositiveMassEvolution}(a,b), the CCE of the intermediate band (at a given $\omega$) evolves smoothly from $\epsilon^{(2)}(\bk-\bq)$ at $t=0$ (blue circle) to $\epsilon^{(1)}(\bk)$ at $t=1$ (orange circle), without ever shrinking to a single point. For the range $|\bQ_i|\leq q_x \leq |\bQ_e|$, Fig.\,\ref{fig: PositiveMassEvolution}(c), the CCEs of $\epsilon^{(1)}(\bk)$ and $\epsilon^{(2)}(\bk-\bq)$ 
intersect at two points, thus all intermediate-band CCEs are constrained to go through those two points, which necessarily forbids the formation of ETPs. In other words, in the limit $\eta\rightarrow0$, $\QPI(\bq, \omega)$ is zero in the three cases analyzed thus far. Eventually, an ETP appears when $q_x = |\bQ_e|$, \textit{i.e.} when the two circles are externally tangent with each other (Fig.\,\ref{fig: PositiveMassEvolution}(d)). In this configuration, the intermediate band CCE evolves from the blue circle at $t=0$ to an ETP at the tangent point, marked by $\otimes$ symbol in Fig.\,\ref{fig: PositiveMassEvolution}(d), and then grows again into the orange circle at $t=1$. Two ETPs appear when $q_x > |\bQ_e|$, Fig.\,\ref{fig: PositiveMassEvolution}(e), where the intermediate band CCE shrinks into an ETP within the blue circle and appears as a distinct ETP within the orange circle before growing into orange band at $t=1$. 

The analysis above already makes it clear why for the particle-particle system the peak at $\bQ_i$ fails to appear. This case corresponds to the configuration shown by Fig.\,\ref{fig: PositiveMassEvolution}(b), where no ETP appears and QPI intensity should be 0. However, although both $|\bq| = |\bQ_e|$ and $|\bq| > |\bQ_e|$ cases have ETPs (Figs.\,\ref{fig: PositiveMassEvolution}(d) and \ref{fig: PositiveMassEvolution}(e)), the calculated $\QPI$ only peaks at $|\bq| = |\bQ_e|$ (Fig.\,\ref{fig: PositiveMassEvolution}(f)). Therefore, the appearance of ETPs is a necessary but not a sufficient condition for QPI.

\begin{figure}
    \centering
    \includegraphics[width=1\linewidth]{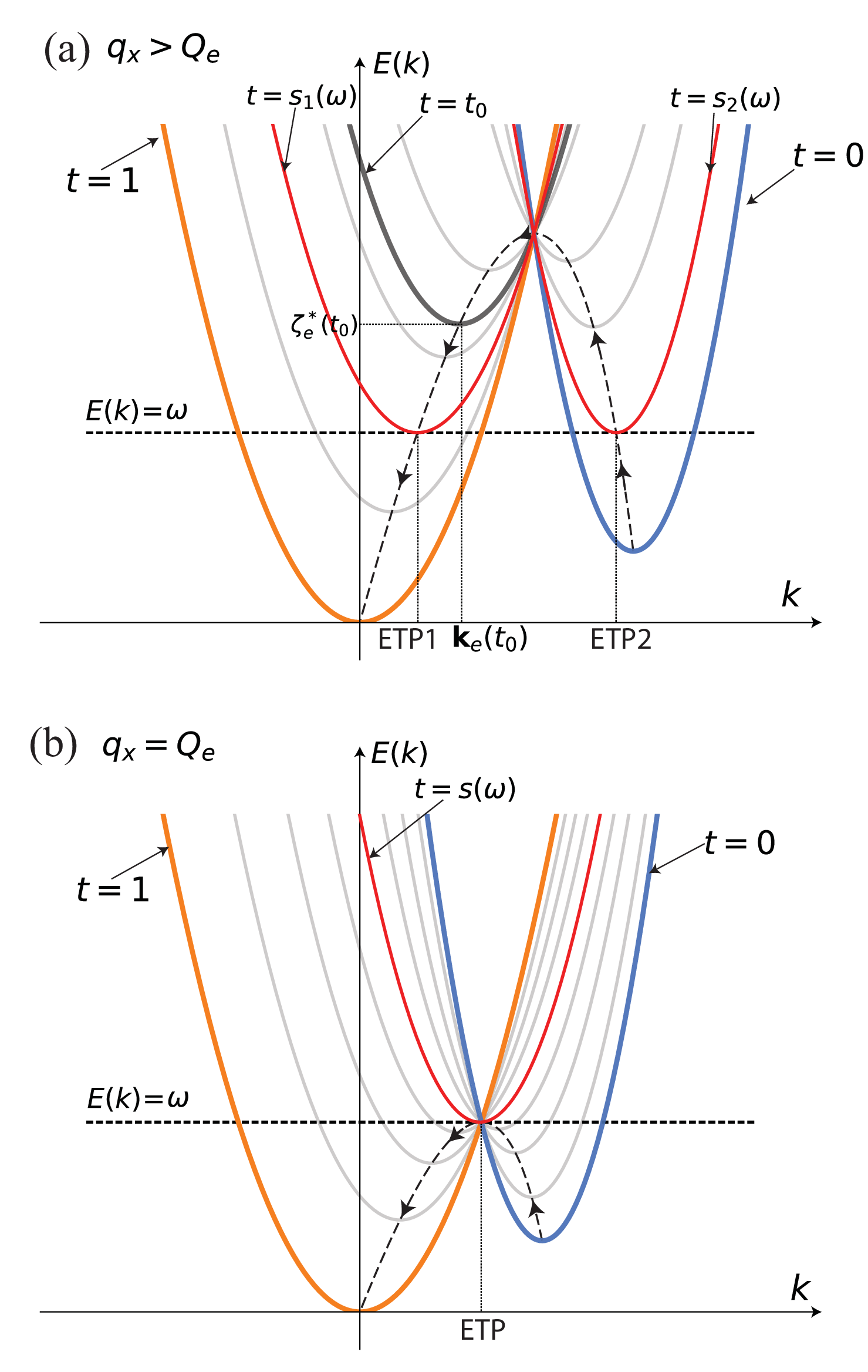}
    \caption{1D illustrations of the intermediate bands evolution when (a) $q>|\bQ_e|$ and (b) $q=|\bQ_e|$.The dashed black curves with arrows in (a) and (b) are traces of the intermediate band extremum when $t$ is changed from 0 to 1, i.e. they are parametric curves of $(\bk_e(t), \epsilon^*_{e}(t))$. In (a), there are two intermediate bands with minima that touches the energy level $\omega$. These two bands, colored by red, corresponds to two solutions $s_1(\omega)$ and $s_2(\omega)$ of the equation $\zeta^*_e(t) = \omega$. When $q=|\bQ_e|$ (b), only one intermediate has its minimum touching $\omega$ and two original bands also intersect at this ETP.}
    \label{fig:Proof1}
\end{figure}

To understand the difference between the ETPs that appear in $q_x=|\bQ_e|$ and $q_x>|\bQ_e|$ situations, we need to further investigate the analytical properties of the integral in Eq.\,(\ref{eq: QPI Integral of DDOS}). Given an intermediate band $\epsilon^*_{t, \bq}(\bk)$, the coordinate of the band extremum is a function of $t$ and is denoted as $\bk_e(t)$ here. The energy value at the extremum is denoted as $\zeta^*_e(t)$:
\begin{equation}
\label{eq: Def Ext Function}
\zeta^*_e(t)=\epsilon^*_{t, \bq}(\bk_e(t))
\end{equation}
With these definitions, the d-DOS for the intermediate band $\epsilon^*_{t, \bq}(\bk)$ is 
\begin{equation}
\frac{\intd g^*_{t, \bq}(\omega)}{\intd \omega} = \frac{\sqrt{m_xm_y}}{2\pi}\delta(\omega-\zeta^*_e(t)) 
\end{equation}
and the QPI intensity $\QPI(\bq, \omega)$ is
\begin{align}
    \QPI(\bq, \omega) & = \frac{1}{2}\int_{0}^{1} \sqrt{m_x(t)m_y(t)}\delta(\zeta^*_e(t)-\omega)\intd t \label{eq: QPI Before Integral}\\
                      & = \frac{1}{2}\sum_{\tau(\omega) \in S}{\sqrt{m_x(\tau(\omega))m_y(\tau(\omega))}}\bigg/{\bigg|\frac{\intd\zeta^*_e}{\intd t}(\tau(\omega))\bigg|} \label{eq: QPI Integrated Form}
\end{align}
where $S$ is the set of roots, $\tau(\omega)$, of the equation $\omega-\zeta^*_e(t)=0$ in the range $0<t<1$. 
Equation (\ref{eq: QPI Integrated Form}) shows that $\QPI$ is dominated by ETPs where $|\intd\zeta^*_e/\intd t|=0$. 
The behavior of $\zeta^*_e$ as a function of $t$ is illustrated by black dashed curves with arrows in Figs.\,\ref{fig:Proof1}(a,b) for the $q_x>|\bQ_e|$ and $q_x=|\bQ_e|$ cases, respectively. Comparing the two figures, one sees that the latter case satisfies $|\intd\zeta^*_e/\intd t|=0$ at the ETP, while the former does not. As shown in the Appendix \ref{Appendix: Zero Derivative Equivalent}, and illustrated in Fig.\,\ref{fig:Proof1}, the condition $|\intd\zeta^*_e/\intd t|=0$ is also equivalent to 
\begin{equation}
\label{eq: Zero Derivative Equivalent}
\epsilon^{(1)}(\bk_e(\tau(\omega))) = \epsilon^{(2)}(\bk_e(\tau(\omega)) - \bq)
\end{equation} 
i.e. it requires the $\bq$ vector to connect points in CCEs, the trivial JDOS restriction. 

We can summarize three necessary conditions for QPI in the intermediate band analysis for a particle-particle system:
\begin{itemize}
    \item[a1] There exist $t_0$ and $\bk_0$ such that $\epsilon^*_{t_0, \bq}(\bk_0) = \omega$.
    \item[a2] $\nabla \epsilon^*_{t_0, \bq}(\bk)|_{\bk=\bk_0} = 0$.
    \item[a3] At $\bk_0$, $\epsilon^{(1)}(\bk_0) = \epsilon^{(2)}(\bk_0 - \bq)$.
\end{itemize}
Next we analyze whether these three conditions are also sufficient to the particle-hole case. We will show that even for the particle-hole case, these three conditions are still true and the geometric meaning of these three conditions directly results in the group velocity selection rule. 

\subsection{Intermediate Bands Analysis of Interband Scattering: Particle-Hole System}
\begin{figure}[!htbp]
    \includegraphics[width=1\linewidth]{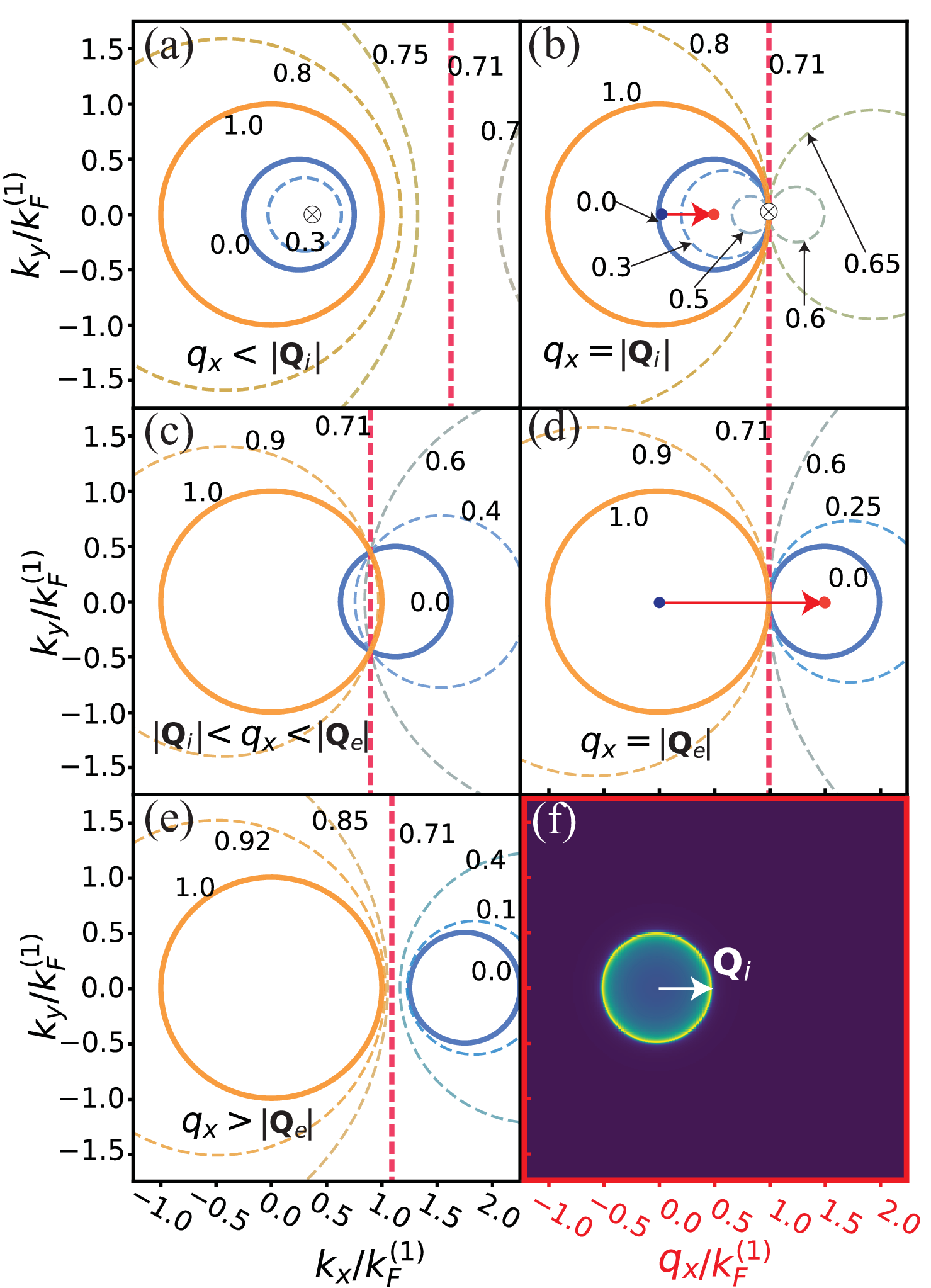}
    \caption{(a)-(e): Evolution of Fermi surfaces (dashed circles) of intermediates with $q_x$ values in different ranges for the particle-hole system given in Fig.\,\ref{fig: Puzzle2D} ($s=-1$). Numbers marked around Fermi surfaces are corresponding $t$ values. ETPs are marked by the symbol $\otimes$ in the figure. Two red arrows in (b) and (d) marks the $\bQ_i$ and $\bQ_e$.(f): Numerical results of QPI response function. (a)-(e) are plots in $\bk$ space (black frame) and (f) is a plot in $\bq$ space (red frame)}
    \label{fig: PosNegMassEvolution}
\end{figure}

The analysis of the interband contributions to $\QPI(\bq, \omega)$ in the particle-hole case is very similar to the particle-particle case, and it can be shown that the conditions for the appearance of interband scattering are the same, \textit{i.e.} (a1-a3). However, for the particle-hole case, the intermediate bands may contain saddle points if the bands are not isotropic. This specific situation is discussed in detail in Appendix \ref{Appendix: Saddle Point}. Here we discuss the simpler case of the isotropic system given by Eqs.\,(\ref{eq: Band1}-\ref{eq: Band2}) with $s=-1$.  whose intermediate bands evolution is shown in Fig.\,\ref{fig: PosNegMassEvolution}.

Comparing the evolution of the intermediate bands between the $s=1$ case, Fig.\,\ref{fig: PositiveMassEvolution}, and the $s=-1$ case, Fig.\,\ref{fig: PosNegMassEvolution}, the most obvious difference is that all the intermediate-band CCEs are circles for the particle-particle system while a vertical-line CCE appears during the evolution in the particle-hole system. This linear CCE appears when 
\begin{equation}
t = t_c = \frac{m_1}{m_1 + m_2},
\end{equation}
a $t$ value for which the intermediate band ceases to have quadratic terms. 
When $q_x < Q_{i}$, Fig.\,\ref{fig: PosNegMassEvolution}(a), the blue circle at $t=0$ first shrinks into an ETP, reappears as a second ETP (on the right region out of frame) at large $k_x$ outside both blue and orange circles, evolves into a constant $k_x$ line at $t = t_c = 0.71$ and bends back until it forms the orange CCE at $t=1$. When $q_x = Q_{i}$, Fig.\,\ref{fig: PosNegMassEvolution}(b), conditions a1-a3 are met, and a peak is observed in the calculated $\QPI$, Fig.\,\ref{fig: PosNegMassEvolution}(f). The remaining three cases, shown in Figs.\,\ref{fig: PosNegMassEvolution}(c-e), depict CCEs evolving from the blue to the orange circles, passing through a constant $k_x$ line but never shrinking into an ETP. As a result, these cases do not produce peaks in $\QPI$ at $Q_{e}$

\subsection{Geometric Meaning of Conditions a1-a3: Opposite Group Velocity Rule \label{sec:geometrical meaning}} 

We can now translate the mathematical conditions a1-a3 into simple geometrical rules. The first and last conditions, a1 and a3, are equivalent to the JDOS condition: the two CCEs intersect at a point $\bk_0$ when they are translated by $\bq$. 
The non-trivial restriction comes from condition a2 requiring the existence of $0\leq t_0 \leq1$ such that:
\begin{equation}
t_0 \nabla \epsilon^{(1)}(\bk_0)  = -(1 - t_0)\nabla\epsilon^{(2)}(\bk_0-\bq)
\end{equation}
First, this implies that the group velocity of two original bands must be anti-parallel at $\bk_0$. Second, given that a1 and a3 also restrict $\bk_0$ to be an intersection point of the two bands, a2 also implies that $\bk_0$ must also be a tangent point of the two CCEs. QPI signals are maximized when $\bq$ connects points in $\bk_i$ space with opposite group velocities, while no signal is observed when the group velocities are parallel -- the opposite group velocity rule.

It is worth noting that although our discussion was restricted to quadratic bands, a similar analysis can be extended to more general band structures. First, note that Eq.\,\ref{eq: QPI Integral of DDOS} is generally true for any band structure. Thus, for $\QPI(\bq,\omega)$ to have a large value, there must exist an intermediate band whose d-DOS diverges at the energy level $\omega$. Given an arbitrary lattice with band structure $\epsilon(\bk)$, its DOS is
\begin{equation}
g(\omega) = \int_{C(\omega)} \frac{\intd \mu}{|\nabla_{\bk}\epsilon(\bk)|}
\end{equation}
where $C(\omega)$ is the CCE at $\omega$. Since a CCE in 2D can be smoothly parametrized as $(k_x(\mu, \omega), k_y(\mu, \omega))$ with $\mu\in[0, 1]$, $g(\omega)$ is always a smooth function except when $C(\omega)$ contains points satisfying $|\nabla_{\bk}\epsilon(\bk)|=0$, \textit{i.e.} a van Hove singularity (VHS). 
% This shows that the QPI can have a peak only if during the intermediates evolution there exit one intermediate band have VHS at $\omega$. 
Applying this VHS criterion to the intermediate bands $\epsilon_{t, \bq}^{*}(\bk)$ implies condition a2, which ultimately leads to the opposite group velocity rule. This generality is also confirmed by numerical evaluations of other band dispersions (Appendix \ref{app:General_band_structures}) \footnote{In more general tight binding models the Green's functions may contain form factors on the numerators. Since such form factors are smooth in $\bk$ space, the argument proposed here are not affected.}.

% In all, the intermediate band analysis allows us to demonstrate that interband QPI calculated via the QPI interference response function is dominated by intersection points of opposite group velocities. Although the discussion above focused on parabolic bands, the opposite group velocities rule is true for general band structures.

\section{INTERMEDIATE BANDS ANALYSIS FOR BOGOLIUBOV QUASIPARTICLE INTERFERENCE \label{sec:BQPI}}

In this section, we apply the intermediate band analysis to interpret QPI in superconductors with anisotropic gaps. The gap anisotropy gives rise to a set of Bogoliubov QPI (BQPI) vectors that can be observed at energies within the superconducting gap. Conversely, QPI measurements allow the momentum structure of the gap function to be extracted.
For example, such BQPI has recently been observed on the Fe-based superconductor $\FeSeS$ near the nematic quantum critical point, where the BQPI measurements revealed a near-nodal gap structure (Fig.\,\ref{fig: FeSeSQPI}(a)), characteristic of a nematic pairing interaction \cite{Nag2024}, which phenomenologically can be expressed as (see Fig.\,\ref{fig: FeSeSQPI}(b)):
\begin{equation}
\label{eq: Gap Function}
\Delta(\theta) = \Delta_s + \Delta_s' \cos^2(2\theta)
\end{equation}
Interestingly, while the dispersion relation of the superconducting band structure is particle-hole symmetric, the experimental BQPI intensity shows a subtle particle-hole asymmetry. Remarkably, a similar particle-hole asymmetry is also captured by the numerical evaluation of $\QPI(\bq,\omega)$. In this section we show that this effect can be understood through the lens of the intermediate band analysis, leading to the conclusion that this asymmetry originates from the Bogoliubov coefficients $u_\bk$ and $v_\bk$.

It is worth noting that the impact of $u_\bk$ and $v_\bk$ on BQPI has been considered in previous studies. In the real space representation, the structure of $u$ and $v$ factors can be captured by measuring the electron density of bound states around point-like defects, which was predicted to take different geometric shapes at positive and negative energies in Zn-doped Bi2212 \cite{Haas2000, Pan2000}. Another type of effect from Bogoliubov coefficients typically occurs in nodal gap superconductors, where $\bq$ vectors connecting two $\bk$ points with same or opposite gap signs may follow different enhancement/suppression rules for various types of scattering centers \cite{Hanaguri2009, Maltseva2009}. This effect arises from terms containing $u_{\bk_i}u_{\bk_f}$ and $v_{\bk_i}v_{\bk_f}$ that appear in the Green's function calculations of QPI, commonly referred to as a coherence factor effect. In this section, we will also refer to the particle-hole asymmetry in the intensity as a coherence factor effect, since this is also caused by $u_{\bk_i}u_{\bk_f}$ and $v_{\bk_i}v_{\bk_f}$ terms. However, the coherence factor effect discussed here differs from previous works as it leads to a particle-hole asymmetry in the BQPI intensity that is not restricted to superconductors with sign-changing gap structures.

\subsection{Particle-Hole Asymmetry in BQPI}

\begin{figure}[h!]
\includegraphics[width=0.95\linewidth]{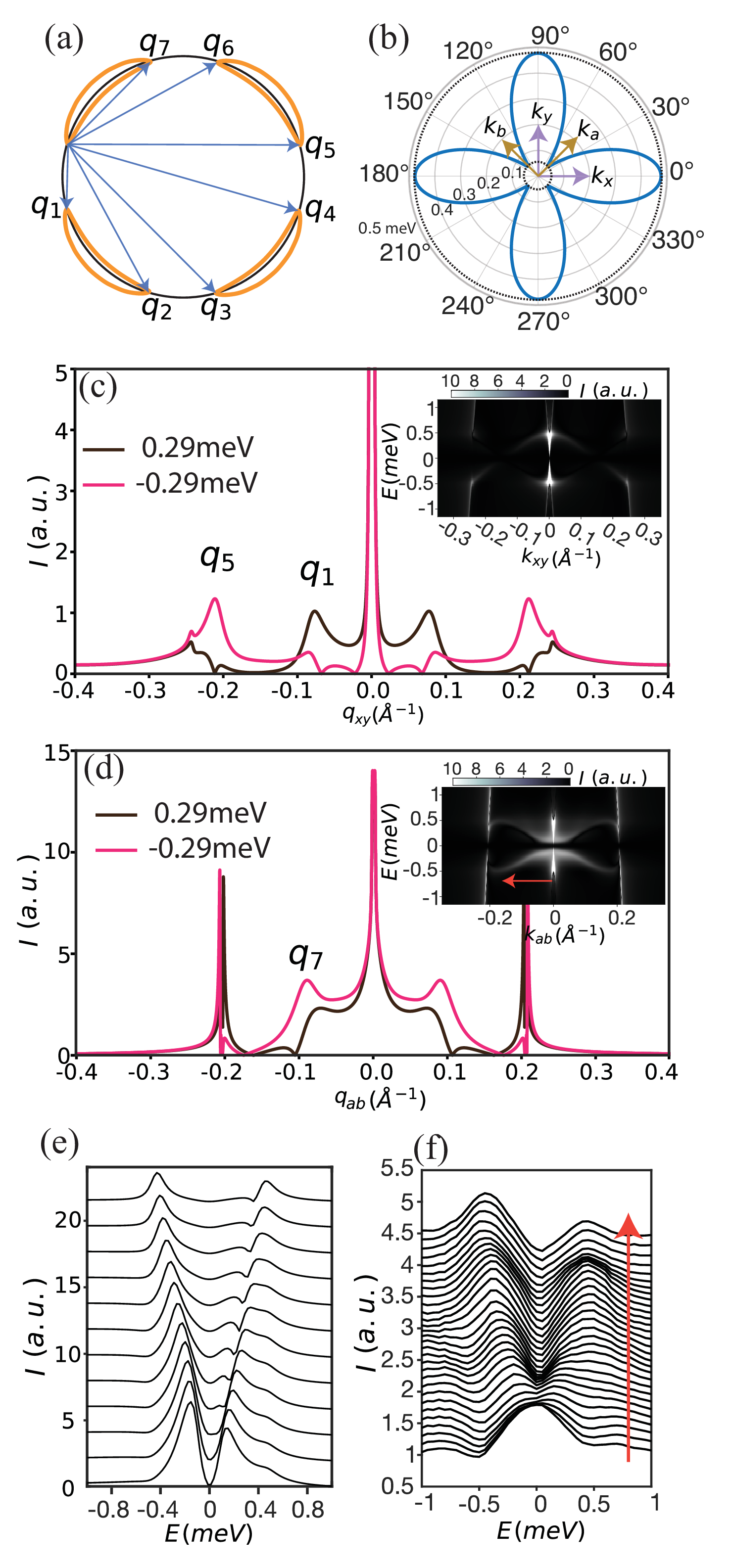}
\caption{(a) A schematic illustration of the \fesesonenine{} superconductor band CCE. The black circle represents the Fermi surface and four closed segments on it are CCEs of superconductor band with the energy between the superconductor gap minimum and maximum. Seven blue arrows mark scattering directions linking high DOS points. (b) Polar plot of the gap function $\Delta(\theta)$ given by Eq.\,(\ref{eq: Gap Function}) with $\Delta_s = 0.0543\mathrm{meV}$ and $\Delta_s' = 0.4228\mathrm{meV}$. This choice of parameters is consistent with our previous work \cite{Nag2024}. (c) QPI intensity along $q_{xy}$ direction at energy $0.29$\,meV. The inserted contour plot is the full line cut QPI map along $q_{xy}$ directions. (d) QPI intensity along $q_{ab}$ direction at energy $0.29$\,meV. The inserted contour plot is the full line cut QPI map along $q_{ab}$ directions. (e)Constant $\bq$ cuts of numerical QPI intensity along $q_{ab}$ direction from $0.0382$\,\AA$^{-1}$ to $0.1377$\,\AA$^{-1}$. (f) Constant $\bq$ cuts of the experimental data in reference \cite{Nag2024} from $0.0384$\,\AA$^{-1}$ to $0.1557$\,\AA$^{-1}$.}
\label{fig: FeSeSQPI}
\end{figure}

In this section, we calculate the BQPI signal for the anisotropic gap structure of \fesesonenine{} and demonstrate that the calculations reveal a particle-hole asymmetry in the intensity of the BQPI vectors. We compare these results to experimental data and show that they are consistent with the STS observations \cite{Nag2024}. Given the gap function $\Delta_\bk$ and the normal state band dispersion $\epsilon_\bk$, the superconductor band structure is \cite{Annett2004}
\begin{equation}
\xi_\bk = \sqrt{\epsilon^2_{\bk} + \Delta^2_{\bk}},
\end{equation}
Within the Nambu-Gor'kov formalism, the QPI intensity $\QPI(\bq, \omega)$ in the superconductor is
\begin{align}
\QPI(\bq, \omega) & = \imag \int\frac{\intd\bk}{(2\pi)^2}G_0(\mathbf{k}, \omega) G_0(\mathbf{k-q},\omega) \nonumber\\
                                           & \pm F_0(\mathbf{k}, \omega) F_0(\mathbf{k-q},\omega) \label{eq: BdG 1st Order Green's Function}
\end{align}
where the sign of the second term depends on the scalar ($-$) or spin-flip ($+$) nature of the scattering \cite{Balatsky2006} and $G_0(\mathbf{k}, \omega)$ and $F_0(\mathbf{k}, \omega)$ are
\begin{align}
G_0(\bk, \omega) = & \frac{u_\bk^2}{\omega - \xi_\bk + i\eta^+} + \frac{v_\bk^2}{\omega + \xi_\bk + i\eta^+} \label{eq: Def G Function}\\
F_0(\bk, \omega) = & \frac{u_\bk v_\bk}{\omega - \xi_\bk + i\eta^+} - \frac{u_\bk v_\bk}{\omega + \xi_\bk + i\eta^+} \label{eq: Def F Function}
\end{align}
where $u_\bk$ and $v_\bk$ are BdG coefficients of the superconductor:
\begin{align}
    u_\bk & = \sqrt{\frac{1}{2}(1 + \frac{\epsilon_\bk^2}{\xi_\bk^2})}\\
    v_\bk & = \sqrt{\frac{1}{2}(1 - \frac{\epsilon_\bk^2}{\xi_\bk^2})}\\
\end{align}

Different from Eq.\,(\ref{eq: Def QPI Convolution}), $\QPI(\bq, \omega)$ in the superconductor case contains an anomalous term $F_0(\mathbf{k}, \omega) F_0(\mathbf{k-q},\omega)$. While this anomalous term plays important rule in the coherence factor effect investigated in previous researches \cite{Hanaguri2009, Maltseva2009}, it is particle-hole symmetric. Therefore, in the rest of this section, we will only focus on the term containing $G_0$ functions, which in this context we call the asymmetric QPI term $\QPI_a(\bq, \omega)$:
\begin{equation}
\label{eq: Def Asymmetric QPI}
\QPI_a(\bq, \omega) = \imag \int\frac{\intd\bk}{(2\pi)^2}G_0(\mathbf{k}, \omega) G_0(\mathbf{k-q},\omega)
\end{equation}

The numerical calculation of $\QPI_a(\bq, \omega)$ is illustrated by Fig.\,\ref{fig: FeSeSQPI}, with $\Delta_\bk$ and $\epsilon_\bk$ determined from experiments \cite{Walker2023, Nag2024}. As illustrated in Fig.\,\ref{fig: FeSeSQPI}(a), the gap function has minima along the $k_a$ and $k_b$ directions, which causes the CCE of the Bogoliubov band to split into four segments for energies between the superconducting gap minimum and maximum. Through the JDOS interpretation, these CCEs result in seven wave vectors $\bq_{1}$-$\bq_{7}$ connecting the points of highest DOS. Figures \ref{fig: FeSeSQPI}(c) and (d) shows line cuts of the calculated $\QPI_a(\bq, \omega)$ along the $q_{a,b}$ direction and the $q_{x,y}$ directions. It shows that the $\bq_5$ and $\bq_7$ peak intensities are weaker at positive energy than at negative energy, whereas the $\bq_3$ peak intensity exhibits the opposite behavior. Among these peaks, the suppression of $\bq_7$ is particularly noteworthy since similar suppression was observed by experiment (see Fig.\,\ref{fig: FeSeSQPI}(e) and (f)). This motivated us to apply the intermediate band analysis to $\QPI_a(\bq, \omega)$ in Eq.\,(\ref{eq: Def Asymmetric QPI}) to understand the observed particle-hole asymmetry. 

\subsection{Intermediate bands analysis for the superconducting state}
\begin{figure}[htbp]
\includegraphics[width=1.0\linewidth]{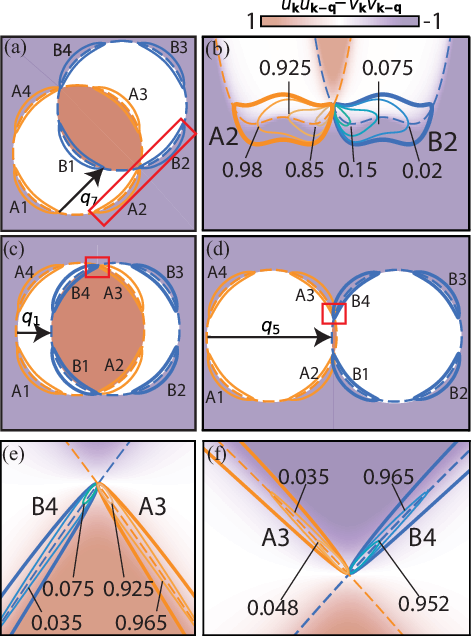}
\caption{(a), (c) and (d) are schematic illustrations of CCE structure of band $\xi_\bk$ and $\xi_{\bk-\bq}$ with $\bq = \bq_7$, $\bq_1$ and $\bq_5$ relatively. (b), (e) and (f) are zoomed-in plots (red boxes in (a), (c) and (d) relatively.) of intermediate band evolution around ETPs. The orientation in (b) is rotated by $45^\circ$ relative to (a) to optimize the representation. These three plots are at energy 0.29 meV.}
\label{fig: FeSeSIntermediate}
\end{figure}
The asymmetric term $\QPI_a$ can be further divided into the particle term ($\QPI_p$), the hole term ($\QPI_h$) and the mixed term ($\QPI_{ph}$) 

% \textcolor{red}{(I added Fig.\,\ref{fig: Fig_IpIh} in appendix to demonstrate that Iph term is much smaller than tow other terms. I personally don't think it's necessary to put this in the paper but if you want I can write a new appendix section to describe this figure or describe this figure in Appendix E.)CAN YOU PLOT THE FOLLOWING THREE TERMS SEPARETELY (FOR US OR APPENDIX), TO DEMONSTRATE THE ASSERTION THAT Iph IS MUCH SMALLER THAN THE OTHERS?}:

\begin{align}
\QPI_h = & \imag \int \frac{u^2_{\bk}}{\omega - \xi_{\bk}} \frac{u^2_{\bk - \bq}}{\omega - \xi_{\bk-\bq}} \frac{\intd^2\bk}{(2\pi)^2} \label{eq: Def Particle Term}\\
\QPI_p = & \imag \int \frac{v^2_{\bk}}{\omega + \xi_{\bk}} \frac{v^2_{\bk - \bq}}{\omega + \xi_{\bk-\bq}}\frac{\intd^2\bk}{(2\pi)^2}\label{eq: Def Hole Term}\\
\QPI_{ph} = & \imag \int \frac{u^2_{\bk}}{\omega - \xi_{\bk}} \frac{v^2_{\bk - \bq}}{\omega + \xi_{\bk-\bq}} + \frac{v^2_{\bk}}{\omega + \xi_{\bk}} \frac{u^2_{\bk - \bq}}{\omega - \xi_{\bk-\bq}} \frac{\intd^2\bk}{(2\pi)^2}\label{eq: Particle-Hole Term}
\end{align}
Here we can further neglect $\QPI_{ph}$ since it only contains first order poles, whose contribution to the final QPI pattern, compared with $\QPI_p$ and $\QPI_h$, can be neglected. 
Since $\xi_\bk$ is positive, $\QPI_h$ and $\QPI_p$ only contain poles at positive and negative energy, respectively.
Thus, $\QPI_p$ only contributes to BQPI peaks appearing below the Fermi level (particle scattering) while $\QPI_h$ only contributes to those above the Fermi level (hole scattering). Given the particle-hole symmetry in the denominators of $\QPI_p$ and $\QPI_h$, the source of the asymmetry is in the numerators, i.e. the coherence factors $u^2_\bk u^2_{\bk - \bq}$ and $v^2_\bk v^2_{\bk - \bq}$. 

The forms of $\QPI_p$ and $\QPI_s$ are similar to the quadratic interband cases discussed earlier, following a two-pole integral form, and the Feynman parametrization can be applied in the same way.
\begin{align}
\QPI_h = & \imag\int \intd t \int \intd \bk\frac{u^2_\bk u^2_{\bk - \bq}}{(\omega - \xi_{t, \bq}^*(\bk) + i\eta^+)^2}\label{eq: Particle Term Intermediate Band}\\
\QPI_p = & \imag\int \intd t \int \intd \bk\frac{v^2_\bk v^2_{\bk - \bq}}{(\omega + \xi_{t, \bq}^*(\bk) + i\eta^+)^2}\label{eq: Hole Term Intermediate Band}
\end{align}
Here, similar to Eq.\,(\ref{eq: Def Intermediate Band}), $\xi_{t, \bq}^*(\bk)$ is the BdG intermediate band:
\begin{equation}
\xi_{t, \bq}^*(\bk) = t \xi(\bk) + (1 - t) \xi(\bk - \bq)
\end{equation}
% Considering the intermediate bands evolution is helpful to qualitatively determine the QPI intensity of $\QPI_p$ and $\QPI_h$. 
However, there are two important differences between the quadratic bands case discussed previously and the $\FeSeS$ superconductor case discussed here. First, unlike the quadratic bands where only ETPs contribute to QPI values, intermediate bands at a range of $t$ values contribute to QPI intensity in the superconductor case. This is because the DOS for the superconducting band $\xi_\bk$ is no longer a step function anymore (\textit{i.e.}, its derivative is finite away from the band extremum) -- see Appendix \ref{Appendix: qt Map} for more information.
Second, in the superconductor case, the intermediate band evolution is masked by the coherence factors $u^2_\bk u^2_{\bk - \bq}$ and $v^2_\bk v^2_{\bk - \bq}$. Given that $\FeSeS$ has a hole band dispersion at the $\Gamma$ point, $u_\bk$ is nearly 1 and $v_\bk$ is nearly 0 at $\bk$ points within the Fermi surface, whereas $u_\bk$ is nearly 0 and $v_\bk$ is nearly 1 at $\bk$ points outside the Fermi surface. When two bands $\xi_\bk$ and $\xi_{\bk-\bq}$ are overlaid, $u^2_\bk u^2_{\bk - \bq}$ has non-vanishing values when both $u^2_\bk$ and $u^2_{\bk - \bq}$ are almost one, corresponding to the intersection area of regions within the two Fermi surfaces. Similarly, $v^2_\bk v^2_{\bk - \bq}$ only has non-vanishing values in regions outside both Fermi surfaces. These two differences cause the particle-hole asymmetry observed in numerical calculations. 

Taking the case of $\bq_7$ as an example, Fig.\,\ref{fig: FeSeSIntermediate}(a) illustrates the CCE structure of $\xi_\bk$ and $\xi_{\bk-\bq_7}$. The four CCE segments for band $\xi_\bk$ are marked as A1-A4, and the four segments for band $\xi_{\bk - \bq}$ are marked as B1-B4. For the $\bq_7$ case, the dominant QPI contribution comes from the A2-B2 and A4-B4 pairs, which are equivalent by symmetry. 
The intermediate band evolution of the A2-B2 pair is shown in Fig.\,\ref{fig: FeSeSIntermediate}(b), where an ETP exists at $t=0.5$. However, as previously discussed, not only the intermediate bands at ETPs but all intermediate bands contribute to the final value of QPI. For a given $t$ in Eqs.\,(\ref{eq: Particle Term Intermediate Band}) and (\ref{eq: Hole Term Intermediate Band}) the contribution of the intermediate band to the QPI can be qualitatively analyzed by considering whether its CCE has significant overlap with the $u^2_\bk u^2_{\bk - \bq}$ and $v^2_\bk v^2_{\bk - \bq}$ masks. As shown in Fig.\,\ref{fig: FeSeSIntermediate}(b), the CCEs for various $t$ show substantial overlap with high-intensity regions of $v^2_\bk v^2_{\bk - \bq}$ (purple shading) but almost no overlap with high-intensity regions of $u^2_\bk u^2_{\bk - \bq}$ (pink shading). This indicates that the QPI intensity at $\bq_7$ is suppressed for particle scattering, i.e. $\bq_7$ is smaller at positive energies relative to negative energies. This also explains the particle-hole asymmetry observed in the experiment and calculation.

A similar argument can be applied to $\bq_1$ and $\bq_5$, as shown in Figs.\,(\ref{fig: FeSeSIntermediate})(c) and (d). Although the dominant contribution of QPI intensity for either $\bq_1$ or $\bq_5$ comes from the A3-B4 and A2-B1 segment pairs, the $u^2_\bk u^2_{\bk - \bq}$ mask has large overlap with these CCEs pairs when $\bq=\bq_1$ (Fig.\,\ref{fig: FeSeSIntermediate}(e)) while the $v^2_\bk v^2_{\bk - \bq}$ mask has a large overlap when $\bq=\bq_5$ (Fig.\,\ref{fig: FeSeSIntermediate}(f)). This means $\bq_1$ is suppressed for particle scattering (negative energies) while the $\bq_5$ is suppressed for hole scattering (positive energies), consistent with the numerical results (Fig.\,\ref{fig: FeSeSQPI}(b)).

\section{Conclusions \label{sec:conclusions}}

QPI measurements have become increasingly important for investigating the electronic structure of solids. A common strategy for interpreting QPI is to examine the contours of constant energy and identify $\bq$ vectors that connect $\bk_f$ and $\bk_i$ points. Such interpretations, based on the JDOS approximation, are relatively straightforward in single-band systems such as the cuprates but become increasingly complex in multiband systems, where a single $\bq$ can correspond to multiple pairs of $\bk_f$ and $\bk_i$, complicating the interpretation of the QPI signal. However, we show that calculations based on the convolution of unperturbed Green’s functions impose additional constraints on the $(\bk_f,\bk_i)$ pairs that contribute to QPI. Specifically, we find that QPI is dominated by pairs with opposite group velocities. This insight corrects the JDOS criteria with an additional rule for identifying dominant QPI signals in experimental observations.

In the process of deriving the opposite group velocity rule, we developed a framework to reformulate the equations for calculating the convolution of Green’s functions in terms of integrals over intermediate bands. This framework proved valuable not only for understanding the generic behavior of inter-band QPI but also for explaining the particle-hole asymmetry observed in the intensity of the Bogoliubov QPI (BQPI) signal in \fesesonenine{}. The intermediate band analysis reveals that this intensity asymmetry arises from the coherence factors intrinsic to the superconducting band structure. To our knowledge, such a coherence factor effect has not been previously discussed.

The existence of this coherence factor effect also has practical implications for experimentalists. As we demonstrated for the hole-like band of \fesesonenine{}, certain BQPI signals (\textit{e.g.}, $\bq_7$) are naturally expected to be more intense at negative energies than at positive energies, while others (\textit{e.g.}, $\bq_1$) are more pronounced at positive energies. These insights provide a practical guide for experimentalists, enabling them to identify the energy ranges where the extraction of BQPI signals achieves the highest fidelity in determining the momentum structure of the superconducting gaps.

Overall, it will be interesting for other experimentalists to apply the group velocity selection rule derived here to interpret their QPI data. These rules can be tested against both past and future experiments to assess how accurately calculations based on the convolution of Green’s functions describe experimental observations. In particular, we suggest focusing on $\bq$ locations where the JDOS predicts signals with parallel group velocities, as our results indicate that such signals should be suppressed. Furthermore, the intermediate band framework introduced here provides a valuable tool for unraveling complex QPI signals. By utilizing this framework, experimentalists may gain deeper insights into the extraction of superconducting gap structures from BQPI data. Notably, the particle-hole asymmetry in BQPI intensity could serve as a practical method for distinguishing between different possible gap structures.

\begin{acknowledgments}
E.H.d.S.N. acknowledges support from the National Science Foundation under grant number DMR-2034345. This work was supported by the Alfred P. Sloan Fellowship (E.H.d.S.N.). A.F.K. was supported by the National Science Foundation under grant no. DMR-1752713.
\end{acknowledgments}

\bibliographystyle{apsrev4-2}
\setcitestyle{notitle}
\bibliography{InterBandPaper}

\appendix
\section{EQUATION {\ref{eq:SOL and DOS relation}}}
\label{Appendix: Ddos Proof}
By definition, the DOS $g(\omega)$ of an arbitrary band $\epsilon(\bk)$ is 
\begin{equation}
g(\omega) = -\frac{1}{\pi}\imag\int\frac{1}{\omega - \epsilon(\bk) + i\eta^+}\frac{\intd \bk^n}{(2\pi)^n}
\end{equation}
Taking derivative on both side gives:
\begin{equation}
\frac{\intd g(\omega)}{\intd \omega} = \frac{1}{\pi}\imag\int\frac{1}{(\omega - \epsilon(\bk) + i\eta^+)^2}\frac{\intd \bk^n}{(2\pi)^n}
\end{equation}
This gives Eq.\;(\ref{eq:SOL and DOS relation}).

\section{EQUATION \ref{eq: Zero Derivative Equivalent}}
\label{Appendix: Zero Derivative Equivalent}
Here we show that $|\intd\zeta^*_e/\intd t|=0$ is equivalent to 
\begin{equation}
\epsilon^{(1)}(\bk_e(s(\omega))) = \epsilon^{(2)}(\bk_e(s(\omega)) - \bq)
\end{equation} 

For two arbitrary bands $\epsilon^{(1)}(\bk)$ and $\epsilon^{(1)}(\bk)$, Eq.\,(\ref{eq: Def Ext Function}) defines:
\begin{align}
\frac{\intd \zeta^*_e(t)}{\intd t} = & \frac{\intd \epsilon^*_{t, \bq}(\bk_e(t))}{\intd t} \\
                                      = & \frac{\intd}{\intd t} [ t \epsilon^{(1)}(\bk_e(t)) + (1 - t) \epsilon^{(2)}(\bk_e(t) - \bq)]\\
                                       = & \epsilon^{(1)}(\bk_e(t)) - \epsilon^{(2)}(\bk_e(t) - \bq) \\
                                         & + \frac{\intd \bk_{e}(t)}{\intd t} \cdot \nabla_{\bk}\epsilon^*_{t, \bq}(\bk_{e}(t))
\end{align}
Since $\bk_{e}(t)$ is already the extremum of the band $\epsilon^*_{t, \bq}(\bk)$, $\nabla_{\bk}\epsilon^*_{t, \bq}(\bk_{e}(t)) = 0$ and 
\begin{equation}
\frac{\intd \zeta^*_e(t)}{\intd t} = \epsilon^{(1)}(\bk_e(t)) - \epsilon^{(2)}(\bk_e(t) - \bq)
\end{equation}
This equation shows that if $\frac{\intd\zeta^*_e(s(\omega))}{\intd t} = 0$, then 
\begin{equation}
\epsilon^{(1)}(\bk_e(s(\omega))) = \epsilon^{(2)}(\bk_e(s(\omega)) - \bq).
\end{equation}

\section{ANALYSIS OF SADDLE POINTS IN THE PARTICLE-HOLE CASE \label{Appendix: Saddle Point}}
One significant difference between a particle-particle system and a particle-hole system is that all intermediate bands of the particle-particle system are parabolic bands while saddle points can exist for a particle-hole system there exist non-physical intermediate bands whose effective masses along $k_x$ and $k_y$ directions has opposite signs. Assuming the following anisotropic particle-hole bands pair:
\begin{align}
    \epsilon^{(1)}(\bk) = & \frac{k_x^2}{2m} + \frac{k_y^2}{2m} + \mu_1\label{eq: InhomoBand1}\\
    \epsilon^{(2)}(\bk) = & -\bigg(\frac{k_x^2}{2M_x} + \frac{k_y^2}{2M_y} + \mu_2\bigg)\label{eq: InhomoBand2}\\
\end{align}
The first band is circular while the second band is elliptic. More general cases where two bands are both elliptic with different orientation angles can always be simplified into this case by translation, scaling and rotation, which only brings a constant multiplier to QPI. Effective masses of the intermediate band in this case are:
\begin{align}
\frac{1}{m_x^*} = & \frac{t}{m} - \frac{(1 - t)}{M_x}\\
\frac{1}{m_y^*} = & \frac{t}{m} - \frac{(1 - t)}{M_y}
\end{align}
It can be seen that $m_x^*$ changes the sign at 
\begin{equation}
    t_x = \frac{m}{m + M_x}
\end{equation}
while $m_y^*$ changes the sign at
\begin{equation}
    t_y = \frac{m}{m + M_y}
\end{equation}
Without loss of generality, we can assume that $M_x>M_y$. Then, the intermediate band $\epsilon^*_{t, \bq}(\bk)$ has hyperbolic CCEs in the range $t_x< t < t_y$. Based on this observation, the intermediate bands evolution can be divided into to types. When $t > t_y$ or $t < t_x$, the intermediate band CCEs are still parabolic and all arguments before for particle-particle system can be applied, i.e. the three conditions a1-a3 are still correct criteria to determine whether the QPI intensity is divergent.

\begin{figure}[htbp]
    \includegraphics[width=1\linewidth]{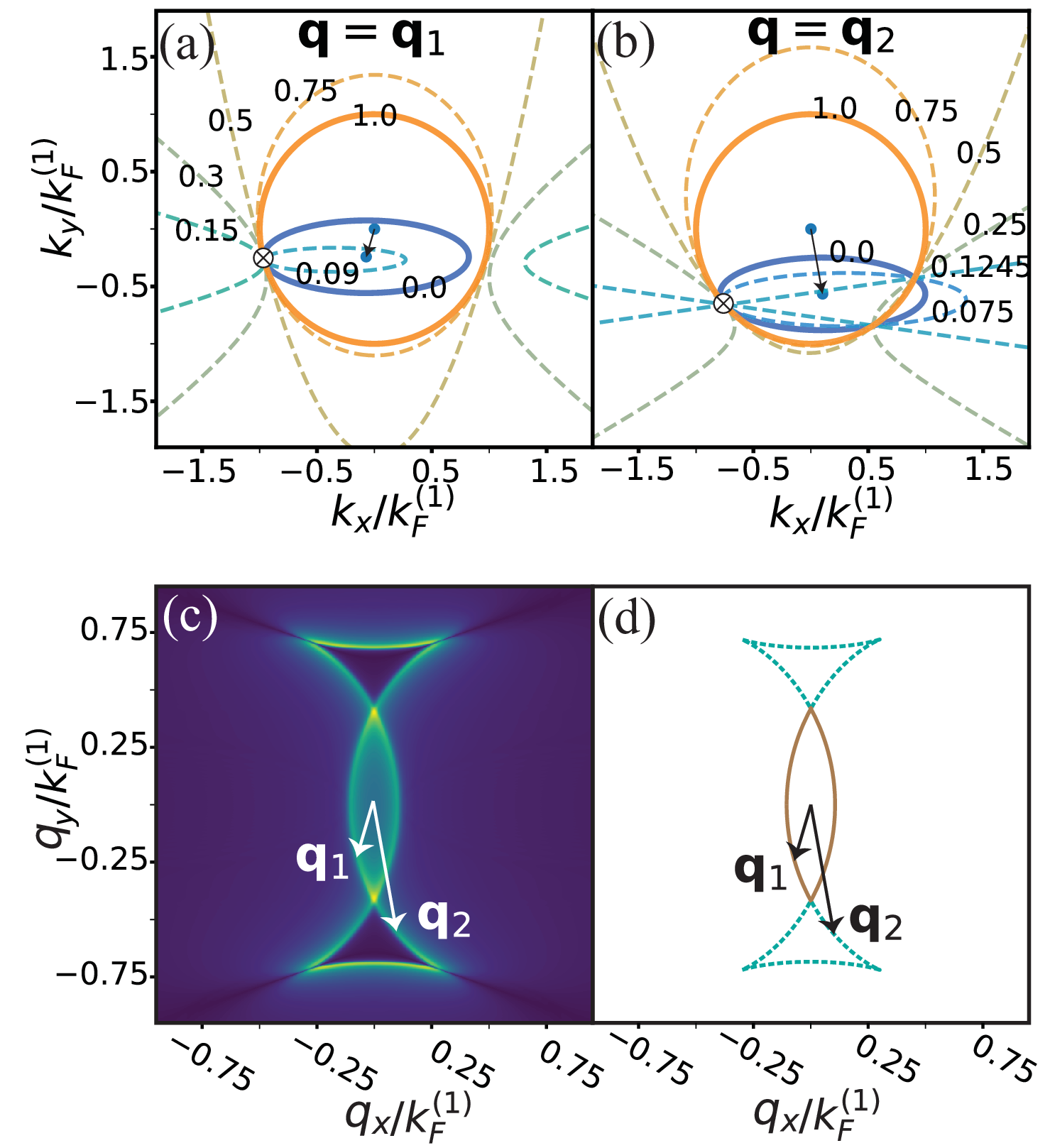}
    \caption{Intermediate band evolution of a particle-hole system with band structure $\epsilon^{(1)}(\bk) = \frac{\bk^2}{2m_1} + \mu_1$ and $\epsilon^{(2)}(\bk) = \frac{k_x^2}{2M_x} + \frac{k_y^2}{2M_y} + \mu_2$ (Eqs.\,(\ref{eq: InhomoBand1}) and (\ref{eq: InhomoBand2})) with $M_x = 8.0 m_1$, $M_y = 0.25 m_1$ and $\mu_2 = 0.1 \mu_1$. (a) The intermediate band evolution at a vector $\bq_1$ with one ETP appears during the evolution. (b) QPI calculated from green's function method. $\bq$ vector marked in A and B are the same for comparison.}
    \label{fig: PosNegInHomo}
\end{figure}

However, when $t_x < t < t_y$, hyperbolic CCEs appear and conditions a1 and a2 corresponds to the case where the saddle point of the band touches the energy level. This is illustrated by Fig.\,\ref{fig: PosNegInHomo}, where saddle point touching appears at $t=0.1425$ and hyperbolic CCE degenerates to two lines crossing at the saddle point. The numerical result for the system illustrated by Fig.\,\ref{fig: PosNegInHomo}(a) is given in Fig.\,\ref{fig: PosNegInHomo}(b). It appears that QPI still diverges at the $\bq$ vector giving saddle point touching. 

To prove that saddle point touching can still give divergent QPI, the d-DOS for an arbitrary hyperbolic band
\begin{equation}
\label{eq: Hyperbolic Band}
\epsilon(k_x, k_y) = \frac{k_x^2}{2m_x} - \frac{k_y^2}{2m_y} + \mu
\end{equation}
should be evaluated first. It is worth noting that the DOS itself is always divergent for a hyperbolic band but by choosing a cutting-off and taking derivative, the d-DOS is finite and is given by:
\begin{equation}
\label{eq: DDOS Hyperbolic Band}
    \frac{\intd g(\omega)}{\intd \omega} = -\frac{1}{2\pi^2}\frac{\sqrt{m_xm_y}}{\omega-\mu}
\end{equation}

\begin{figure}
    \centering
    \includegraphics{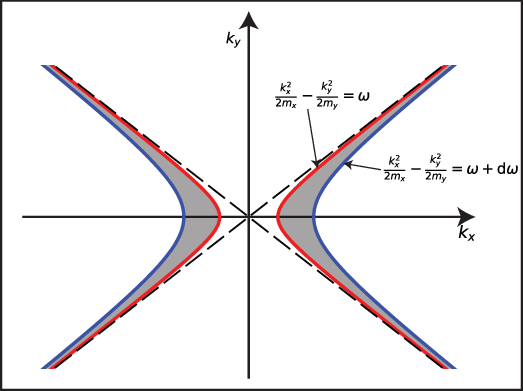}
    \caption{CCEs of the hyperbolic band at $\omega$ and $\omega+\intd\omega$. By definition, the gray region with two CCEs are proportional to the DOS at $\omega$.}
    \label{fig: Proof2}
\end{figure}

To prove Eq.\,\ref{eq: DDOS Hyperbolic Band}, considering the CCEs of the hyperbolic band given by Eq.\,\ref{eq: Hyperbolic Band} at energy level $\omega (\omega>\mu)$:
\begin{equation}
k_x = \pm\sqrt{2m_x \left( \frac{k_y^2}{2m_y} + \omega - \mu \right)}
\end{equation}
Assuming a small deviation of energy $\delta \omega$, the area $\delta A$ between the CCEs at $\omega$ and $\omega + \delta \omega$ is (gray regions in Fig.\,\ref{fig: Proof2})
\begin{align}
    \delta A &= 2\int_{-\infty}^{+\infty}\bigg(\sqrt{2m_x(\frac{k_y^2}{2my} + \omega-\mu+\delta\omega)} \\
            &-\sqrt{2m_x(\frac{k_y^2}{2my} + \omega-\mu)}\bigg)\intd k_y\\
            &=4\delta \omega\int_{0}^{+\infty}\frac{\sqrt{m_xm_y}}{\sqrt{y^2+\omega-\mu}}\intd y
\end{align}
So the DOS per area for a hyperbolic band is:
\begin{align}
    g(\omega) &=\frac{\delta A}{\delta \omega} \frac{1}{(2 \pi)^2}\nonumber\\
              &=\frac{1}{\pi^2}\int_{0}^{+\infty}\frac{\sqrt{m_xm_y}}{\sqrt{y^2+\omega-\mu}}\intd y
\end{align}
The integral in the last step is divergent. However, assuming a cutting-off $\Lambda$, the DOS $g(\omega)$ takes finite value and it's derivative is 
\begin{align}
    \frac{\intd g(\omega)}{\intd \omega} &= -\frac{1}{2\pi^2}\int_{0}^{\Lambda}\frac{\sqrt{m_xm_y}}{(y^2+\omega-\mu)^{3/2}}\intd y\\
                    &= -\frac{1}{2\pi^2}\frac{\sqrt{m_xm_y}}{\omega-\mu}(\Lambda\to\infty)
\end{align}

It is worth noting this result is consistent with the formula for the d-DOS of a normal elliptic band (Eq.\,\ref{eq:Quadratic DDOS}). By the definition of d-DOS given by Eq.\,(\ref{eq:SOL and DOS relation}), the function $\frac{\intd g(\omega)}{\intd \omega}$ should be firstly evaluated at the complex value energy $\omega + i \eta^+$ and then let $\eta^+\to0$. This means that $-\frac{1}{2\pi^2}\frac{\sqrt{m_xm_y}}{\omega-\mu}$ is the real part of a more general function:
\begin{equation}
\label{eq: General DOS}
    \frac{\intd G(\omega)}{\intd \omega} = -\frac{1}{2\pi^2}\frac{\sqrt{m_xm_y}}{\omega-\mu+i\eta^+}
\end{equation}
According to Sokhotski–Plemelj formula, the imaginary part of $G(\omega)$ is just the d-DOS for a band with same effective mass signs.

According to Eq.\,(\ref{eq: DDOS Hyperbolic Band}), the d-DOS for a band with hyperbolic CCEs also diverges at the saddle points, similar to the divergence at the extremum in a parabolic band. This means that the necessary condition for divergent QPI in a particle-hole system is the appearance of either ETPs or saddle point during the intermediate band evolution. Denoting the band saddle point coordinate as $\bk_e(t)$ and the band saddle point value as $\zeta^*_e(t)$, the QPI intensity contributed by the hyperbolic band, according to Eq.\,(\ref{eq: QPI Integral of DDOS}) and Eq.\,(\ref{eq: DDOS Hyperbolic Band}), is:
\begin{equation}
     \label{eq: Integral DDOS Hyperbolic Band}
     \int_{t_y}^{t_x}\frac{\intd g^*_{t, \bq}(\omega)}{\intd \omega}\intd t = -\mathrm{P}\int_{t_y}^{t_x}\frac{1}{2\pi^2}\frac{\sqrt{m_x(t)m_y(t)}}{\omega-\zeta^*_e(t)}\intd t
\end{equation}
If the saddle point touching happens at $(t_x<t_0<t_y)$, i.e. $\omega=\zeta^*_e(t_0)$, there is one singularity at the integral of Eq.\,(\ref{eq: Integral DDOS Hyperbolic Band}). However, similar to the particle-particle case, this singularity can only cause a divergence after the integration when $\omega - \zeta^*_e(t)$ cease to have the first order expansion at $t_0$ since the principal value for the integral with a first order pole is still finite. This means that when saddle points touch the energy level during the intermediate band evolution, the sufficient condition for a divergent QPI is still $|\intd\zeta^*_e/\intd t|=0$. In conclusion, the condition for a intermediate band with opposite effective mass sign to cause a divergence in QPI is exactly the same as for the parabolic case. Thus, the conditions a1-a3 for particle-particle systems can be equally applied to particle-hole systems. 

The similarity between the ETP and saddle point touchings discussed above is illustrated in Fig.\,\ref{fig: PosNegInHomo}, which considers the QPI between an isotropic and an anisotropic band. In Fig.\,\ref{fig: PosNegInHomo}(a) when $\bq=\bq_1$, there is one ETP at the tangent point of the ellipse and circle, which causes a divergent QPI feature in the $\eta\rightarrow0$ limit -- similar to the $q_x=|\bQ_i|$ case in Fig.\,\ref{fig: PosNegMassEvolution}(b). Interestingly, the manifold of QPI vectors that occur from these ETPs, brown curve in Fig.\,\ref{fig: PosNegInHomo}(d), does not constitute the full QPI set obtained from the numerical evaluation of $\QPI$, Fig.\,\ref{fig: PosNegInHomo}(c). The remaining set occurs, for example, when $\bq=\bq_2$, Fig.\,\ref{fig: PosNegInHomo}(b), which is the case when there are two intersections and one tangent point between the CCEs of the two original bands. Since intermediate band CCEs are forced to go through these three points, the CCE cannot evolve into a single point anymore. Instead, the intermediate-band CCE evolves firstly into a CCE composed of two lines crossing at the tangent point at $t=0.1245$, then into a hyperbola (e.g.\,$t=0.5$) and then into the orange CCE at $t=1$. This tangent point marked by $\otimes$ in Fig.\,\ref{fig: PosNegInHomo}(b) is the saddle point of the intermediate band and also causes divergence in the QPI pattern. As a result, QPI peaks at both $\bq_1$ and $\bq_2$, as confirmed by the numerical evaluation (Fig.\,\ref{fig: PosNegInHomo}(c)).

The opposite group velocity rule discussed in Sec.\,\ref{sec:geometrical meaning} also provides a method to determine the curve of divergence of QPI in $\bk$ space for two general quadratic bands. If we denote the CCEs of the two quadratic bands as $S^{(1)}$ and $S^{(2)}$, then the curve of divergence can be construced through two steps. First, translate $S^{(2)}$ until it is tangent with $S^{(1)}$. This tangent should be external for a particle-particle system and internal for a particle-hole system to satisfy the opposite group velocity rule. Then, keep these two curves to be tangential and move $S^{(2)}$ to make it go around $S^{(1)}$ without rotation. The center of $S^{(2)}$ traces a curve and this is the curve of divergence in QPI. The curve of divergence predicted by the opposite group velocity rule is consistent with numerical calculations and previous analytical results. Figures \ref{fig: PosNegInHomo}(c) and (d) shows the numerical Green's function result and the curve of divergence predicted by the selection rule and they matches with each other. It is worth mentioning that in this specific system, the curve of divergence has two segments corresponds to two cases where the divergence is caused by ETPs (brown solid line) or saddle touching points (green dashed line). For another, the opposite group velocity rule predicts that a single quadratic band $\epsilon(\bk) = \frac{k_x^2}{2m_x} + \frac{k_y^2}{2m_y} - \mu$ has the curve of divergence $\frac{q_x^2}{2m_x} + \frac{q_y^2}{2m_y} = 4(\omega + \mu)$. The analytical result derived in Ref.\,\cite{Capriotti2003} shows the same pole structure.

\begin{figure*}[!t]
    \centering
    \includegraphics[width=0.95\linewidth]{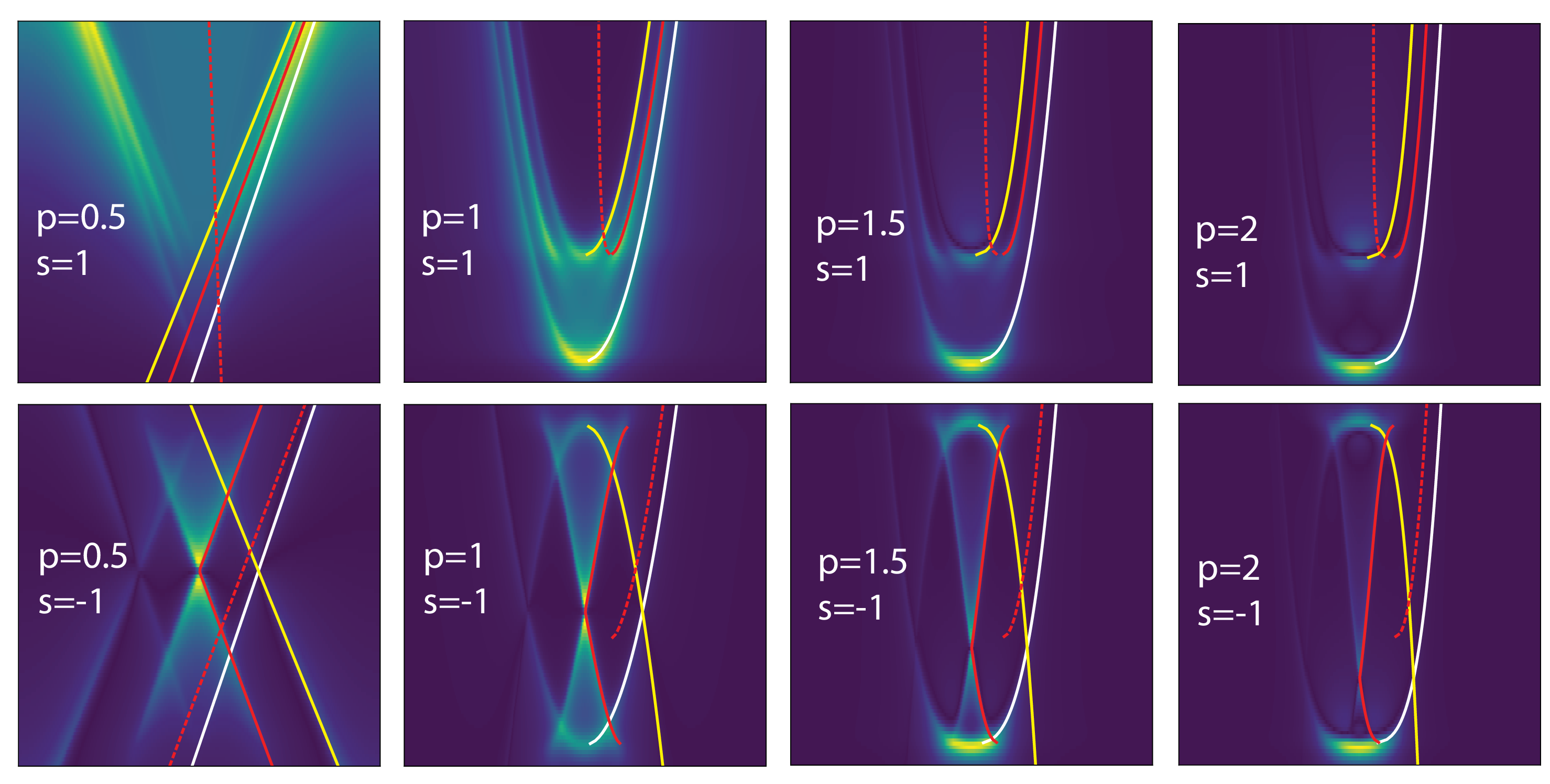}
    \caption{QPI simulated by numerical Green's function method. The white and yellow solid lines mark QPI corresponding to intraband $\bq$ vectors from the band 1 and band 2. The red solid lines mark interband QPI $\bq$ vectors connecting two point with anti-parallel group velocities while the red dashed lines mark interband QPI $\bq$ vectors connecting two point with parallel group velocities.}
    \label{fig: GeneralBands}
\end{figure*}

\section{SELECTION RULE FOR GENERAL BAND DISPERSIONS \label{app:General_band_structures}}
In main text we proved the anti-parallel group velocity rule for the pure quadratic band case. In a real material, the band dispersion is more complicated and can have higher order corrections. To check whether the opposite group rule is true for more general bands dispersion, we numerically calculated QPI for linear (Dirac), cubic and fourth order bands. The two bands system in the calculation are given by
\begin{align}
    \label{eq: GeneralBand1} \epsilon^{(1)}(\bk) = & \frac{(k_x^2 + k_y^2)^p}{2m_1} - \mu_1\\
    \label{eq: GeneralBand2} \epsilon^{(2)}(\bk) = & s\frac{(k_x^2 + k_y^2)^p}{2m_2} - s\mu_2
\end{align}
To simplify the parameterization, we fix the Fermi level at 0 and choose $\mu_1$ as the energy unit and $k_F^{(1)}$ as the $\bk$ scale unit, which uniquely determines the unitless form of the first band:
\begin{equation}
\label{eq: UnitlessGeneralBand1}
\frac{\epsilon^{(1)}(\bk)}{\mu_1} = \bigg(\big(\frac{k_x}{k_F^{(1)}}\big)^2 + \big(\frac{k_y}{k_F^{(1)}}\big)^2 \bigg)^p - 1
\end{equation}
The expression of the second band can be further determined by two more unitless parameters $\alpha=\frac{\mu_2}{\mu_1}$ and $\beta=\frac{k_F^{(2)}}{k_F^{(1)}}$, which gives
\begin{equation}
\label{eq: UnitlessGeneralBand2}
\frac{\epsilon^{(2)}(\bk)}{\mu_1} = s\alpha \bigg(\frac{\big(\frac{k_x}{k_F^{(1)}}\big)^2 + \big(\frac{k_y}{k_F^{(1)}}\big)^2}{\beta^2} \bigg)^p - s\alpha
\end{equation}
In our calculation, we choose $\alpha=0.5$ and $\beta=0.6$ and calculation results are illustrated in Fig.\,\ref{fig: GeneralBands}, which shows that the selection rule all appears to be true for different dispersion relations. 

% \begin{figure}
%     \centering
%     \includegraphics[width=0.95\linewidth]{Fig_IpIh.pdf}
%     \caption{Numerical result of $I_p$, $I_h$ and $I_ph$ given by Eqs.\,(\ref{eq: Def Particle Term}), (\ref{eq: Def Hole Term}) and (\ref{eq: Particle-Hole Term})}
%     \label{fig: Fig_IpIh}
% \end{figure}

\section{$\bq-t$ MAP ANAYSIS FOR $\bq_7$}
\label{Appendix: qt Map}
\begin{figure}[!h]
    \centering
    \includegraphics[width=0.95\linewidth]{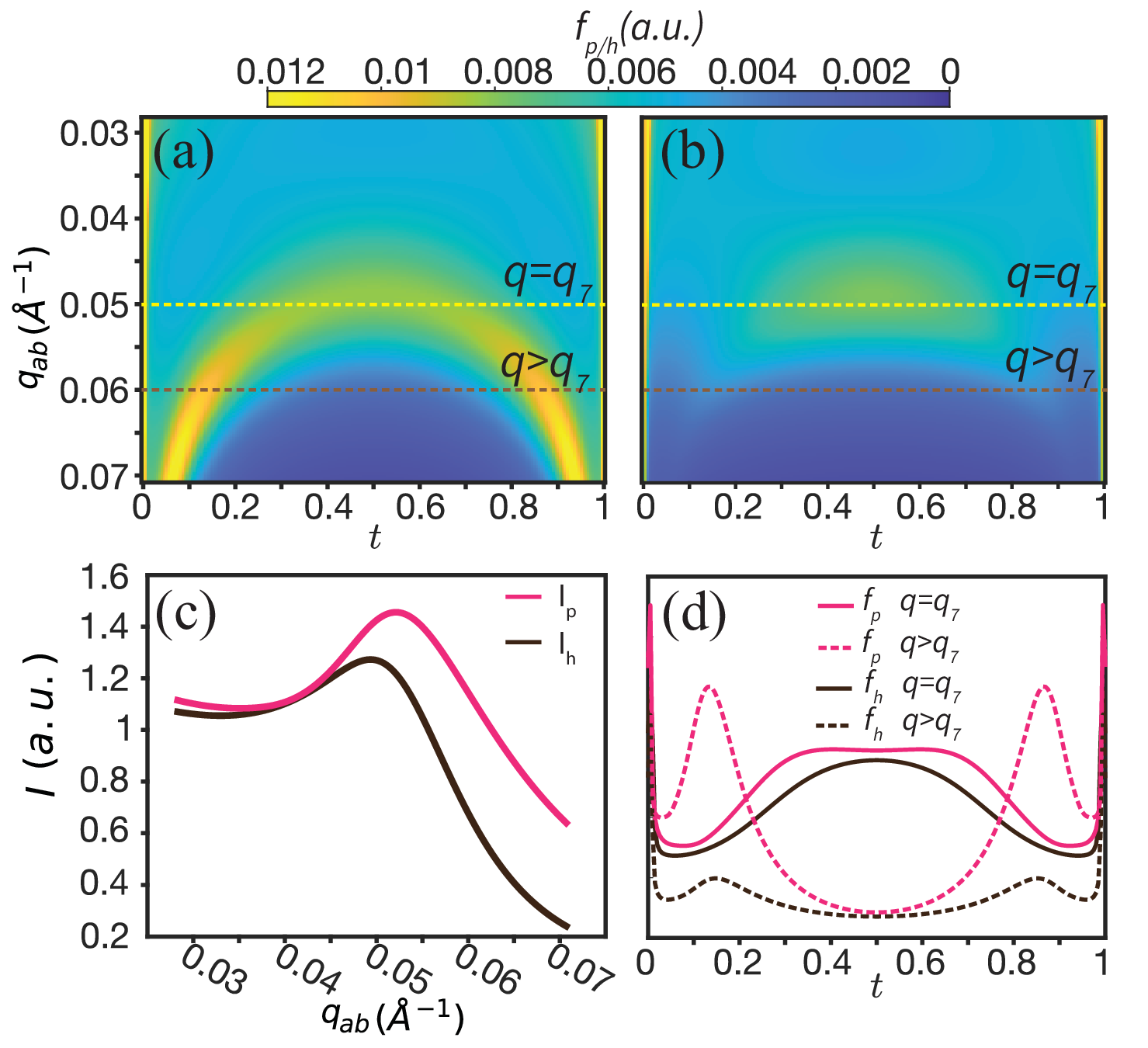}
    \caption{(a) The $\bq-t$ maps of $f_p(t, \bq, \omega)$ at $\omega=0.15\mathrm{meV}$. (b) The $\bq-t$ maps of $f_h(t, \bq, \omega)$ at $\omega=-0.15\mathrm{meV}$. $\bq$ in (a) and (b) are along the $q_{ab}$ direction. The yellow dashed lines in (a) and (b) mark the $|\bq_7|$ value ($0.05\mathrm{\AA^{-1}}$). (c) $I_p$ and $I_h$ intensity obtained by integrating along the $t$ axis. (d) Line cuts along $q=0.05\mathrm{\AA^{-1}}$ (yellow dashed lines in (a) and (b)) and $q=0.06\mathrm{\AA^{-1}}$ (brown lines in (a) and (b)). In this calculation, the $\eta^+$ in Green's functions are chosen to be $0.03\textrm{meV}$. This value is different from $0.07\mathrm{meV}$ used in calculations illustrated by Fig.\,\ref{fig: FeSeSQPI}.}
    \label{fig: qtMAP}
\end{figure}

To visualize the intermediate band evolution for the BQPI analysis, we also plot here the $\bq-t$ map. A $\bq-t$ map corresponds to the value of $\bk$ integrals in different $x$ and $|\bq|$ values, i.e. the intensity plot of Fig. \ref{fig: qtMAP}(a) and (b) represents following two functions
\begin{align}
f_h(t, \bq, \omega) = & \imag\int \intd \bk\frac{u^2_\bk u^2_{\bk - \bq}}{(\omega - \xi_{t, \bq}^*(\bk) + i\eta^+)^2}\label{eq: Particle Term Intermediate Band qt map}\\
f_p(t, \bq, \omega) = & \imag \int \intd \bk\frac{v^2_\bk v^2_{\bk - \bq}}{(\omega + \xi_{t, \bq}^*(\bk) + i\eta^+)^2}\label{eq: Hole Term Intermediate Band qt map}
\end{align}
with $\omega = 0.15\mathrm{meV}$ for $f_h$ and $\omega = -0.15\mathrm{meV}$ for $f_p$ and $\bq = (q_{ab}, q_{ab})$. As is shown in Fig.\,\ref{fig: qtMAP}(a, b), when $q_{ab} \geq |\bq_7|$ (below the yellow dashed line), a bright arc corresponding to ETPs appears. This ETP arc divides the $\bq-t$ map into two regions. Within the arc, the values of $f_h(t, \bq, \omega)$ and $f_p(t, \bq, \omega)$ is exactly 0 since there is no CCE at $|\omega| = 0.15\mathrm{meV}$. Outside the arc, the values are finite because for a non-quadratic band, all intermediate band CCEs can contribute to the QPI intensity. By integrating the $\bq-t$ maps along the $t$ axis, the QPI intensity can be obtained, which, as is illustrated by Fig.\,\ref{fig: qtMAP}(c), also shows the particle-hole asymmetry discussed in main text. 

The effect of $u_{\bk}u_{\bk-\bq}$ and $v_{\bk}v_{\bk-\bq}$ masks can be observed in $\bq-t$ maps. When $q_{ab} = |\bq_7|$ (yellow dashed lines in Figs.\,\ref{fig: qtMAP}(a, b)), the intensity of $f_{p/h}(t, \bq, \omega)$ is large around $t=0.5$, which corresponds to ETPs. However, as is mentioned in the main text, intermediate bands away from ETPs may also finitely contribute to QPI intensity and their contributions are effect by $u_{\bk}u_{\bk-\bq}$ and $v_{\bk}v_{\bk-\bq}$ masks. This can be seen in the line cut of intensity along $q_{ab} = |\bq_7|$ (Fig.\,\ref{fig: qtMAP}(d)), in which around $x=0.5$, the intensity of $f_p(t, \bq, \omega)$ is always larger than $f_h(t, \bq, \omega)$ due to larger overlapping of $v_{\bk}v_{\bk-\bq}$ mask with the intermediate band evolution (Fig.\,\ref{fig: FeSeSIntermediate}(b)). Similarly, when $q_{ab} > |\bq_7|$ (brown dashed lines in Figs.\,\ref{fig: qtMAP}(a, b)), the intermediate band evolution still have large overlapping with $v_{\bk}v_{\bk-\bq}$ mask and cause larger intensity for $f_p(t, \bq, \omega)$, as is shown by the line cuts in Fig.\,\ref{fig: qtMAP}. Especially, two peaks induced by ETPs at $t=0.15$ and $0.85$ are considerably suppressed for $f_p(t, \bq, \omega)$. This is consistent with the calculation result illustrated by Fig.\,\ref{fig: FeSeSQPI}(d), in which the QPI intensity is still larger at negative energy even when $q_{ab}>|\bq_7|$.

There is one final comment about the line cuts in Fig.\,\ref{fig: qtMAP}(d). It can be seen that the peak corresponding to ETPs is higher at $q_{ab} > |\bq_7|$ rather then $q_{ab} = |\bq_7|$. This does not mean that QPI intensity, i.e., the values obtained by integrating along $t$ axis, should be larger at $q_{ab} > |\bq_7|$. This is because the peak at $q_{ab} = |\bq_7|$ is wider than those at $q_{ab} > |\bq_7|$, which compensates the low peak height and results in a QPI peak at $\bq_7$.

% \textcolor{red}{This is very confusing to me. This paragrpah does not clearly explain how to look at this figure.}
% Two important features appears in the $\mathbf{q}-x$ map. Firstly, the intensity is high at the red dash lines (traces of ETPs) for $\QPI_h$ but is suppressed for $\QPI_p$, which is the effect of the $u_{\bk}u_{\bk-\bq}$ and $v_{\bk}v_{\bk-\bq}$ masks. Secondly, the value is still finite in the region below the red dash lines  but is 0 above the red dashed line. This is because for $\FeSeS$, d-DOS is not a delta function at band extremum anymore and whenever the intermediate CCE appears at the energy level we are looking at, these is a finite value contribution of this intermediate band for total QPI intensity. However, when $|\bq|>|\bq_7|$, the intermediate band at some $t$ value is fully above or below the energy level and in this case the value of d-DOS is exactly 0.
% \textcolor{red}{IN THE MAIN TEXT, THIS APPENDIX IS CALLED TO GIVE MORE DETAILS IN SUPPORT OF: ``intermediate bands at a range of t values contribute to QPI intensity in the superconductor case.'' I DON'T SEE HOW THIS PLOT SUPPORTS THIS. IF I LOOK AT THE HORIZONTAL LINE FOR q7 AND I IMAGINE INTEGRATING OVER t I WOULD GUESS TO OBTAIN THE SAME VALUE FOR Ip AND Ih. \textbf{ARE THE LABELS FOR Ih and Ip INVERTED?}}

\end{document}